\documentclass[12pt,oneside]{article}
\usepackage{amsmath,amssymb,amsfonts,amsthm}
\usepackage{color}

\newcommand{\T}{{\cal T}}

\newcommand{\prof}{\noindent \textit{\textbf{Proof.\:\:}}}
\newcommand{\tm}{\T M}

\def\o#1{\overline{#1}}
\def\pa{\partial}

\newcommand\overcirc[1]{\raisebox{10pt}{\tiny$\circ$}{\kern-6.5pt}\mbox{$#1$}}
\newcommand\undersym[2]{\raisebox{-6pt}{\tiny$#2$}{\kern-5pt}\mbox{$#1$}}

\def\Section#1{\vspace{30truept}\addtocounter{section}{1}\setcounter{thm}{0}\setcounter{equation}{0}
{\noindent\Large\bf\arabic{section}.~~#1}\par \vspace{12pt}}

\newtheorem{thm}{Theorem}[section]
\newtheorem{cor}[thm]{Corollary}
\newtheorem{lem}[thm]{Lemma}
\newtheorem{prop}[thm]{Proposition}
\newtheorem{defn}[thm]{Definition}

\newtheorem{rem}[thm]{Remark}

\numberwithin{equation}{section}
\begin{document}
\title{\bf{Geometry of Parallelizable Manifolds\\ in the Context of\\ Generalized Lagrange Spaces}}
\author{M. I. Wanas$^\dagger$,
  N. L. Youssef$^{\,\ddagger}$ and  A. M.  Sid-Ahmed$^{\ddagger}$}
\date{}
%\thanks{\it Department of Mathematics, etc}
%\pagestyle{fancy}

\maketitle                     % Produces the title.
\vspace{-1.13cm}
\begin{center}
{$\dagger$ Department of Astronomy, Faculty of Science, Cairo
University}
\end{center}
\vspace{-0.8cm}
\begin{center}
 wanas@frcu.eun.eg
 \end{center}
\vspace{-0.5cm}
\begin{center}
{$\ddagger$ Department of Mathematics, Faculty of Science, Cairo
University}\end{center} \vspace{-0.8cm}
\begin{center}
{nyoussef@frcu.eun.eg, \ \  amrs@mailer.eun.eg }
\end{center}

\vspace{1cm} \maketitle
\smallskip

\noindent{\bf Abstract.} In this paper, we deal with a
generalization of the geometry\linebreak of parallelizable
manifolds, or the absolute parallelism (AP-) geometry, in the
context of generalized Lagrange spaces. All geometric objects
defined in this geometry are not only functions of the positional
argument $x$, but also depend on the directional argument $y$. In
other words, instead of dealing with geometric objects defined on
the manifold $M$, as in the case of classical AP-geometry, we are
dealing with geometric objects in the pullback bundle $\pi^{-1}(TM)$
(the pullback of the tangent bundle $TM$ by $\,\pi: T
M\longrightarrow M$). Many new geometric objects, which have no
counterpart in the classical AP-geometry, emerge in this more
general context. We refer to such a geometry as generalized
AP-geometry (GAP-geometry). In analogy to AP-geometry, we define a
$d$-connection in $\pi^{-1}(TM)$ having remarkable properties, which
we call the canonical $d$-connection, in terms of the unique
torsion-free Riemannian $d$-connection. In\linebreak addition to
these two $d$-connections, two more $d$-connections are defined, the
dual and the symmetric $d$-connections. Our space, therefore, admits
twelve curvature tensors (corresponding to the four defined
$d$-connections), three of which vanish identically. Simple formulae
for the nine non-vanishing curvatures tensors are obtained, in terms
of the torsion tensors of the canonical $d$-connection. The
different $W$-tensors admitted by the space are also calculated. All
contractions of the $h$- and $v$-curvature tensors and the
$W$-tensors are derived. Second rank symmetric and skew-symmetric
tensors, which prove useful in physical applications, are singled
out. This paper, however, is not an end in itself, but rather the
beginning of a research direction. The physical interpretation of
the geometric objects in the GAP-space that have no counterpart in
the classical AP-space will be further investigated in forthcoming
papers. \footnote {This paper was presented in \lq\lq\,The
International Conference on Finsler Extensions of Relativity
 Theory\rq\rq \ held at Cairo, Egypt, November 4-10,
2006. ArXiv Number: 0704.2001.}

\bigskip
\medskip\noindent{\bf Keywords:}  Parallelizable manifold, Generalized Lagrange space, AP-geometry, GAP-geometry,
Canonical $d$-connection, $W$-tensor.

\bigskip
\medskip\noindent{\bf 2000 AMS Subject Classification.\/} 53B40, 53A40, 53B50.

\newpage
%%%%%%%%%%%%%%%%%%%%%%%%%%%%%%%%%%%%%%%%%%%%% Section 1: Introduction %%%%%%%%%%%%%%%%%%%%%%%%%%%%%%%%%%%%%%%%%%%%%%%%%%%%%%%%%%%

\Section{Introduction} The geometry of parallelizable  manifolds or
the absolute parallelism geometry (AP-geometry) (\cite{FI},
\cite{HP}, \cite{b}, \cite{AMR}) has many advantages in comparison
to Riemannian geometry. Unlike Riemannian geometry, which has ten
degrees of freedom (corresponding to the metric components for $n =
4$), AP-geometry has sixteen degrees of freedom (corresponding to
the number of components of the four vector fields defining the
parallelization). This makes AP-geometry a potential candidate for
describing physical phenomena other than gravity. Moreover, as
opposed to Riemannian geometry, which admits only one symmetric
linear connection, AP-geometry admits at least four natural
(built-in) linear connections, two of which are non-symmetric and
three of which have non-vanishing curvature tensors. Last, but not
least, associated with an AP-space, there is a Riemannian structure
defined in a natural way. Thus, AP-geometry contains within its
geometrical structure all the mathematical machinary of Riemannian
geometry. Accordingly, a comparison between the results obtained in
the context of AP-geometry and general relativity, which is based on
Riemannian geometry, can be carried out.
\par
In this paper, we study AP-geometry in the wider context of a
generalized\linebreak Lagrange space (\cite{RR}, \cite{CC},
\cite{L}, \cite{NN}). All geometric objects defined in this space
are not only functions of the positional argument $x$, but also
depend on the directional argument $y$. In other words, instead of
dealing with geometric objects defined on the manifold $M$, as in
the case of classical AP-space, we are dealing with geometric
objects in the pullback bundle $\pi^{-1}(TM)$ (the pullback of the
tangent bundle $TM$ by the projection $\,\pi: T M\longrightarrow M$)
\cite{N}. Many new geometric objects, which have no counterpart in
the classical AP-space, emerge in this more general context. We
refer to such a space as a $d$-parallelizable manifold or a
generalized absolute parallelism space (GAP-space).
\par
The paper is organized in the following manner. In section 2,
following the introduction, we give a brief account of the basic
concepts and definitions that will be needed in the sequel,
introducing the notion of a non-linear connection
$N^{\alpha}_{\mu}$. In section 3, we consider an $n$-dimensional
$d$-parallelizable manifold $M$ (\cite{DF}, \cite{L}) on which we
define a metric in terms of the $n$ independent $\pi$-vector fields
\ $\undersym{\lambda}{i}$ defining the parallelization on
$\pi^{-1}(TM)$. Thus, our parallelizable manifold becomes a
generalized Lagrange space, which is a generalization of the
classical AP-space. We then define the canonical $d$-connection $D$,
relative to which the $h$- and $v$-covariant derivatives of the
vector fields $\,\undersym{\lambda}{i}$ vanish. We end this section
with a comparison between the classical AP-space and the GAP-space.
In section 4, commutation formulae are recalled and some identities
obtained. We then introduce, in analogy to the AP-space, two other
$d$-connections: the dual $d$-connection and the symmetric
$d$-connection. The nine non-vanishing curvature tensors,
corresponding to the dual, symmetric and Riemannian $d$-connections
are then calculated, expressed in terms of the torsion tensors of
the canonical $d$-connection.  In section 5, a summary of the
fundamental symmetric and skew symmetric second rank tensors is
given, together with the symmetric second rank tensors of zero
trace. In section 6, all possible contractions of the $h$- and $v$-
curvature tensors are obtained and the contracted curvature tensors
are expressed in terms of the fundamental tensors given in section
5. In section 7, we study the different $W$-tensors corresponding to
the different $d$-connections defined in the space, again expressed
in terms of the torsion tensors of the canonical $d$-connection.
Contractions of the different $W$-tensors and the relations between
them are then derived. Finally, we end this paper by some concluding
remarks.

%%%%%%%%%%%%%%%%%%%%%%%%%%%%%%%%%%%%%%%%%%%%%%%% Section 2: Fundamental preliminaries %%%%%%%%%%%%%%%%%%%%%%%%%%%%%%%%%%%%%%%%%%%%%

\Section{Fundamental Preliminaries}

 Let $M$ be a differential manifold of dimension $n$ of class $C^{\infty}$. Let $\pi:TM\to M$
be its tangent bundle. If $(U, \ x^{\mu})$ is a local chart on $M$,
then ($\pi^{-1}(U), \ (x^{\mu}, y^{\mu}$)) is the corresponding
local chart on $TM$. The coordinate transformation law on $TM$ is
given by:
$$x^{\mu'} = x^{\mu'}(x^{\nu}), \ \ \ y^{\mu'} = p^{\mu'}_{\nu} y^{\nu},$$
where $p^{\mu'}_{\nu} = \frac{\partial {x^{\mu'}}}{{\partial
x^{\nu}}}$ and $\text {det} (p^{\mu'}_{\nu})\neq 0.$

\begin{defn} A non-linear connection $N$ on $TM$ is a system of $n^{2}$ functions
$N^{\alpha}_{\beta}(x, \ y)$ defined on every local chart $\pi^{
-1}(U)$ of TM which have the transformation law
\begin{equation}N^{\alpha'}_{\beta'} = p^{\alpha'}_{\alpha} p^{\beta}_{\beta'} N^{\alpha}_{\beta} +
p^{\alpha'}_{\epsilon}
p^{\epsilon}_{\beta'\sigma'}y^{\sigma'},\end{equation} where
$p^{\epsilon}_{\beta'\sigma'} = \frac{\partial
{p^{\epsilon}_{\beta'}}} {\partial x^{\sigma'}} = \frac{\partial^{2}
x^{\epsilon}}{\partial x^{\beta'}\partial x^{\sigma'}}.$
\end{defn}
The non-linear connection $N$ leads to the direct sum decomposition
$$T_{u}(TM) = H_{u}(TM)\oplus V_{u}(TM), \ \ \forall\  u\in \tm = TM\setminus \{0\},$$
where $H_u(TM)$ is the {\it horizontal} space at $u$ associated with
$N$ supplementary to the {\it vertical} space $V_{u}(TM)$. If
$\delta_{\mu}: = \partial_{\mu} - N^{\alpha}_{\mu}\dot
{\partial}_{\alpha}$, where $\partial_{\mu}: =
\frac{\partial}{\partial x^{\mu}}$, $\dot {\partial_{\mu}}: =
\frac{\partial}{\partial y^{\mu}}$, then $(\dot{\partial_{\mu}})$ is
the natural basis of $V_u(TM)$ and  $(\delta_{\mu})$ is the natural
basis of $H_{u}(TM)$ adapted to $N$.

\begin{defn} A distinguished connection (d-connection) on $M$ is a triplet
$D = (N^{\alpha}_{\mu}, \Gamma^{\alpha}_{\mu\nu},
C^{\alpha}_{\mu\nu}$), where $N^{\alpha}_{\mu}(x, y)$ is a
non-linear connection on $TM$ and $\Gamma^{\alpha}_{\mu\nu}(x, y)$
and $C^{\alpha}_{\mu\nu}(x, y)$ transform according to the following
laws:
\begin{equation}\Gamma^{\alpha'}_{\mu'\nu'} = p^{\alpha'}_{\alpha}
p^{\mu}_{\mu'}p^{\nu}_{\nu'}\Gamma^{\alpha}_{\mu\nu} +
p^{\alpha'}_{\epsilon} p^{\epsilon}_{\mu'\nu'},\end{equation}
\begin{equation}C^{\alpha'}_{\mu'\nu'} = p^{\alpha'}_{\alpha}
p^{\mu}_{\mu'}p^{\nu}_{\nu'}C^{\alpha}_{\mu\nu}.\end{equation} In
other words, $\Gamma^{\alpha}_{\mu\nu}$ transform as the
coefficients of a linear connection, whereas $C^{\alpha}_{\mu\nu}$
transform as the components of a tensor.
\end{defn}
\begin{defn} The horizontal (h-) and vertical (v-) covariant derivatives
with respect to the d-connection $D$ (of a tensor field
$A^{\alpha}_{\mu}$) are defined respectively by:
\begin{equation}A^{\alpha}_{{\mu}|{\nu}}: =
\delta_{\nu}A^{\alpha}_{\mu} +
A^{\epsilon}_{\mu}\Gamma^{\alpha}_{\epsilon\nu} -
A^{\alpha}_{\epsilon} \Gamma^{\epsilon}_{\mu\nu};
\end{equation}

\begin{equation}
A^{\alpha}_{{\mu}||\nu} := \dot{\partial}_{\nu}A^{\alpha}_{\mu} +
A^{\epsilon}_{\mu}C^{\alpha}_{\epsilon\nu} - A^{\alpha}_{\epsilon}
C^{\epsilon}_{\mu\nu}.
\end{equation}
\end{defn}

\begin{defn} A symmetric and non-degenerate tensor field $g_{\mu\nu}(x, y)$
of type (0, 2) is called a generalized Lagrange metric on the
manifold $M$. The pair $(M, \ g)$ is called a generalized Lagrange
space.
\end{defn}

\begin{defn} Let $(M, g)$ be a generalized Lagrange space equipped
with a non-linear connection $N^{\alpha}_{\mu}$. Then a d
-connection $D = (N^{\alpha}_{\mu}, \Gamma^{\alpha}_{\mu,\nu},
C^{\alpha}_{\mu\nu})$ is said to be metrical with respect to $g$ if
\begin{equation}g_{\mu\nu|\alpha} = 0, \ \ \ g_{\mu\nu||\alpha} = 0.
\end{equation}
\end{defn}
The following remarkable result was proved by R. Miron \cite{R}. It
guarantees the existence of a unique {\it torsion-free} metrical
$d$-connection on any generalized Lagrange space equipped with a
non-linear connection. More precisely:

\begin{thm} Let $(M, g)$ be a generalized Lagrange space.
Let $N^{\alpha}_{\mu}$ be a given non-linear connection on $TM$.
Then there exists a unique metrical $d$-connection $ \ \overcirc{D}
= (N^{\alpha}_{\mu}, \ \overcirc {\Gamma}^{\alpha}_{\mu\nu}, \
\overcirc {C}^{\alpha}_{\mu\nu})$ such that \
$\overcirc{\Lambda}^{\alpha}_{\mu\nu}: =
 \ \overcirc{\Gamma}^{\alpha}_{\mu\nu} - \ \overcirc{\Gamma}^{\alpha}_{\nu\mu} = 0$
and \ $\overcirc{T}^{\alpha}_{\mu\nu}: = \
\overcirc{C}^{\alpha}_{\mu\nu} -
 \ \overcirc{C}^{\alpha}_{\nu\mu} = 0$. This
$d$-connection is given by $N^{\alpha}_{\mu}$ and the generalized
Christoffel symbols:
\begin{equation}\overcirc{\Gamma}^{\alpha}_{\mu\nu} = \frac{1}{2}
g^{\alpha\epsilon}(\delta_{\mu} g_{\nu\epsilon} + \delta_{\nu}
g_{\mu\epsilon} - \delta_{\epsilon} g_{\mu\nu}),\end{equation}
\begin{equation}\overcirc{C}^{\alpha}_{\mu\nu} = \frac{1}{2}
g^{\alpha\epsilon}(\dot{\partial}_{\mu} g_{\nu\epsilon} +
\dot{\partial}_{\nu} g_{\mu\epsilon} - \dot{\partial}_{\epsilon}
g_{\mu\nu}).\end{equation}
\end{thm}
This connection will be referred to as the Riemannian
$d$-connection.

%%%%%%%%%%%%%%%%%%%%%%%%%%%%%%%%%%%%%%%%%%% Section 3: d-parallelizable manifolds %%%%%%%%%%%%%%%%%%%%%%%%%%%%%%%%%%%%%%%%%%%%%%%%%%%%

\Section{$d$-Parallelizable manifolds (GAP-spaces)} The Riemannian
$d$-connection mentioned in Theorem 2.6 plays the key role in our
generalization of the AP-space, which, as will be revealed, appears
natural. However, it is to be noted that the close resemblance of
the two spaces is deceptive; as they are similar in form. However,
the extra degrees of freedom in the generalized AP-space makes it
richer in  content and different in its geometric structure (see
Remark 3.6).
\par
We start with the concept of $d$-parallelizable manifolds.
\vspace{-5pt}
\begin{defn} An $n$-dimensional manifold $M$ is called d-parallelizable,
or generalized absolute parallelism space (GAP-space), if  the
pull-back bundle $\pi^{-1}(TM)$ admits  $n$ global linearly
independent sections ($\pi$-vector fields)\
$\undersym{\lambda}{i}(x, \ y)$, $i = 1, ..., n$.
\end{defn}
\vspace{-5pt} If \ $\undersym{\lambda}{i} = ( \
\undersym{\lambda}{i}^{\alpha})$, $\alpha = 1, ..., n$,
then\vspace{-5pt}
\begin{equation}\undersym{\lambda}{i}^{\alpha} \ \undersym{\lambda}{i}_{\beta} = \delta^{\alpha}_{\beta}, \ \ \
\undersym{\lambda}{i}^{\alpha} \ \undersym{\lambda}{j}_{\alpha} =
\delta_{ij},\vspace{-5pt}\end{equation} where $( \
\undersym{\lambda}{i}\,_{\alpha})$ denotes the inverse of the matrix
$( \ \undersym{\lambda}{i}^{\alpha})$. \vspace{7pt}
\par
Einstein summation convention is applied on both Latin (mesh)
indices and Greek (world) indices, where all Latin indices are
written in a lower position.
\par
In the sequel, we will simply use the symbol $\lambda$ (without a
mesh index) to denote any one of the vector fields \
$\undersym{\lambda}{i}$ $(i = 1,..., n)$ and in most cases, when
mesh indices appear they will be in pairs, meaning summation.
\par
We shall often use the expression GAP-space (resp. GAP-geometry)
instead of $d$-parallelizable manifold (resp. geometry of
$d$-parallelizable manifolds) for its typographical simplicity.
\pagebreak
\begin{thm} A GAP-space is a generalized Lagrange space.
\end{thm}
In fact, the covariant tensor field $g_{\mu\nu}(x, y)$ of order $2$
given by\vspace{-5pt}
\begin{equation}g_{\mu\nu}(x, y): =
\undersym{\lambda}{i}_{\mu} \
\undersym{\lambda}{i}_{\nu},\vspace{-5pt}\end{equation} defines a
metric in the pull-back bundle $\pi^{-1}(TM)$ with inverse given
by\vspace{-5pt}
\begin{equation}g^{\mu\nu}(x, y) = \undersym{\lambda}{i}^{\mu} \ \undersym{\lambda}{i}^{\nu}
\vspace{-5pt}\end{equation}
\par
Assume that $M$ is a GAP-space equipped with a non-linear connection
$N^{\alpha}_{\mu}$. By Theorem 2.6, there exists  on $(M,g)$ a
unique torsion-free metrical $d$-connection \ $\overcirc{D} =
(N^{\alpha}_{\mu}, \ \overcirc {\Gamma}^{\alpha}_{\mu\nu}, \
\overcirc {C}^{\alpha}_{\mu\nu})$ (the Riemannian $d$-connection).
%\par
We define another $d$-connection $D = (N^{\alpha}_{\mu},
\Gamma^{\alpha}_{\mu\nu}, C^{\alpha}_{\mu\nu})$ in terms of
$\,\overcirc{D}$ by:
\begin{equation}\Gamma^{\alpha}_{\mu\nu}: = \ \overcirc {\Gamma}^{\alpha}_{\mu\nu} +
\undersym{\lambda}{i}^{\alpha} \ \undersym{\lambda}{i}_{\mu{o\atop
|}\nu},\end{equation} \vspace{- 0.6 cm}
\begin{equation}
C^{\alpha}_{\mu\nu}: = \ \overcirc {C}^{\alpha}_{\mu\nu} +
\undersym{\lambda}{i}^{\alpha} \ \undersym{\lambda}{i}_{\mu{o\atop
||}\nu}.\end{equation} Here, \lq\lq\,${o\atop |}$\,\rq\rq\, and
\lq\lq\,${o\atop ||}$\,\rq\rq\, denote the $h$- and $v$-covariant
derivatives with respect to the Riemannian $d$-connection
$\,\overcirc{D}$. If \lq\lq\,$|$\,\rq\rq\, and
\lq\lq\,$||$\,\rq\rq\, denote the $h$- and $v$-covariant derivatives
with respect to the $d$-connection $D$, then\vspace{-5pt}
\begin{equation}{\lambda}^{\alpha}\!\, _{|\mu} = 0, \ \ \
{\lambda}^{\alpha}\!\,_{||\mu} = 0.\vspace{-5pt}\end{equation} This
can be shown as follows:  ${\lambda}^{\alpha}\!\, _{|\mu} =
\delta_{\mu} {\lambda}^{\alpha} +
{\lambda}^{\epsilon}\Gamma^{\alpha}_{\epsilon\mu} =
\delta_{\mu}{\lambda}^{\alpha} + {\lambda}^{\epsilon}( \
\overcirc{\Gamma}^{\alpha}_{\epsilon\mu} +
\undersym{\lambda}{j}^{\alpha} \
\undersym{\lambda}{j}_{\epsilon{o\atop |}\mu}) =
(\delta_{\mu}{\lambda}^{\alpha} + {\lambda}^{\epsilon} \
\overcirc{\Gamma}^{\alpha}_{\epsilon\mu}) -
\undersym{\lambda}{j}^{\alpha}\,\!_{{o\atop |}\mu}( \
\undersym{\lambda}{i}^{\epsilon} \
\undersym{\lambda}{j}_{\epsilon})=0$. In exactly the same way, it
can be shown that ${\lambda}^{\alpha}\!\,_{||\mu} = 0$.  Hence, we
obtain the following
\begin{thm} Let $(M, \ \undersym{\lambda}{i}(x, y))$ be a
GAP-space equipped with  a non-linear connection $N^{\alpha}_{\mu}$.
There exists a unique d-connection $D = (N^{\alpha}_{\mu},
\Gamma^{\alpha}_{\mu\nu}, C^{\alpha}_{\mu\nu})$,
 such that \
${\lambda}^{\alpha}\!\,_{|\mu} = {\lambda}^{\alpha}\!\,_{||\mu} =
0$. This connection is given by $N^{\alpha}_{\beta}$, (3.4) and
(3.5).  Consequently, D is metrical: $g_{\mu\nu|\sigma} =
g_{\mu\nu||\sigma} = 0$.
\end{thm}
This connection will be referred to as the canonical $d$-connection.
\vspace{5pt}
\par
It is to be noted that relations (3.6) are in accordance with the
classical AP-geometry in which the covariant derivative of the
vector fields $\lambda$ with respect to the canonical connection
$\Gamma^{\alpha}_{\mu\nu} = \ \undersym{\lambda}{i}^{\alpha}
(\pa_{\nu}\,\,\undersym{\lambda}{i}_{\mu})$  vanishes \cite{AMR}.

\begin{thm} Let $(M, \ \undersym{\lambda}{i}(x, y))$ be a
d-parallelizable manifold equipped with a non-linear connection
$N^{\alpha}_{\mu}$. The canonical d-connection $D =
(N^{\alpha}_{\mu}, \Gamma^{\alpha}_{\mu\nu}, C^{\alpha}_{\mu\nu})$
is explicitly expressed in terms of \ ${\lambda}$ in the form
\begin{equation}\Gamma^{\alpha}_{\mu\nu} = \undersym{\lambda}{i}^{\alpha}
(\delta_{\nu} \ \undersym{\lambda}{i}_{\mu}), \ \ \
C^{\alpha}_{\mu\nu} =
\undersym{\lambda}{i}^{\alpha}(\dot{\partial}_{\nu} \
\undersym{\lambda}{i}_{\mu}).
\end{equation}
\end{thm}

%\begin{proof}
\prof  Since \ ${\lambda}^{\alpha}\!\, _{|\nu} = 0$, we have
$\delta_{\nu}{\lambda}^{\alpha} = -
{\lambda}^{\epsilon}\,\Gamma^{\alpha}_{\epsilon\nu}$. Multiplying
both sides by ${\lambda}_{\mu}$, taking into account the fact that \
$\undersym{\lambda}{i}^{\alpha} \ \undersym{\lambda}{i}_{\mu} =
\delta^{\alpha}_{\mu}$, we get $\Gamma^{\alpha}_{\mu\nu} = - \
\undersym{\lambda}{i}_{\mu}(\delta_{\nu} \
\undersym{\lambda}{i}^{\alpha}) = \
\undersym{\lambda}{i}^{\alpha}(\delta_{\nu} \
\undersym{\lambda}{i}_{\mu}).$ The proof of the second relation is
exactly similar and we omit it. \ \ $\Box$
%\end{proof}
\vspace{5pt}
\par It is to be noted that the components of the canonical
$d$-connection are similar in form to the components of the
canonical connection in the classical AP-context \cite{AMR}, noting
that $\partial_{\nu}$ is replaced by $\delta_{\nu}$ (for the
$h$-counterpart) and by $\dot{\partial}_{\nu}$ (for the
$v$-counterpart) respectively (See Table 1). The above expressions
for the canonical connection seem therefore like a natural
generalization of the classical  AP case. \vspace{5pt}
\par
 By (3.4) and (3.5), in view of the above theorem, we
have the following\vspace{-0.2cm}
\begin{cor} The Reimannian d-connection \ $\overcirc{D} = (N^{\alpha}_{\mu}, \
\overcirc{\Gamma}^{\alpha}_{\mu\nu}, \
\overcirc{C}^{\alpha}_{\mu\nu})$ is explicitely expressed in terms
of \ $\undersym{\lambda}{i}$ in the form
\begin{equation}\overcirc{\Gamma}^{\alpha}_{\mu\nu} =
\ \undersym{\lambda}{i}^{\alpha}(\delta_{\nu} \
\undersym{\lambda}{i}_{\mu} -
 \ \undersym{\lambda}{i}_{\mu{o\atop |}\nu}), \ \ \ \overcirc{C}^{\alpha}_{\mu\nu} =
\ \undersym{\lambda}{i}^{\alpha}(\dot {\partial}_{\nu} \
\undersym{\lambda}{i}_{\mu} -
 \ \undersym{\lambda}{i}_{\mu{o\atop ||}\nu}).\end{equation}
\end{cor}
\begin{rem} {\em As a result of the dependence of $\,\lambda\,$ on the velocity
vector $y$, the $n^{3}$ functions \
$\undersym{\lambda}{i}^{\alpha}(\partial_{\nu} \
\undersym{\lambda}{i}_{\mu})$, as opposed to the classical AP-space,
do not transform as the coefficients of a linear connection, but
transform according to the rule\vspace{-5pt}
\begin{equation}\undersym{\lambda}{i}^{\alpha'}(\partial_{\nu'} \ \undersym
{\lambda}{i}_{\mu'}) = p^{\alpha'}_{\alpha} p^{\mu}_{\mu'}
p^{\nu}_{\nu'} \ \undersym{\lambda}{i}^{\alpha}(\partial_{\nu} \
\undersym{\lambda}{i}_{\mu}) + p^{\alpha'}_{\epsilon}
p^{\epsilon}_{\mu'\nu'} + p^{\alpha'}_{\alpha} p^{\mu}_{\mu'}
p^{\nu}_{\nu'\epsilon'}
y^{\epsilon'}C^{\alpha}_{\mu\nu}.\vspace{-5pt}
\end{equation} Similarily, it can be
shown that, in general, tensors in the context of the classical
AP-space do not transform like tensors in the wider context of the
GAP-space; their dependence on the velocity vector $y$ spoils their
tensor character.
 In other words, tensors in the classical
AP-context do not necessarily behave like tensors {\it when they are
regarded as functions of position $x$ and velocity vector $y$.} This
means that though the classical AP-space and the GAP-space appear
similar in form, they differ radically in their geometric
structures.}
\end{rem}

We now introduce some tensors that will prove useful later on.
Let\vspace{-5pt}
\begin{equation}\gamma^{\alpha}_{\mu\nu}: = \
\undersym{\lambda}{i}^{\alpha} \ \undersym{\lambda}{i}_{\mu{o\atop
|}\nu} = \Gamma^{\alpha}_{\mu\nu} -  \
\overcirc{\Gamma}^{\alpha}_{\mu\nu}, \ \ \ \ G^{\alpha}_{\mu\nu}: =
\ \undersym{\lambda}{i}^{\alpha} \ \undersym{\lambda}{i}_{\mu{o\atop
||}\nu} = C^{\alpha}_{\mu\nu} -
\overcirc{C}^{\alpha}_{\mu\nu}.\vspace{-5pt}
\end{equation}
In analogy to the AP-space, we refer to $\gamma^{\alpha}_{\mu\nu}$
and $G^{\alpha}_{\mu\nu}$ as the $h$- and $v$-contortion tensors
respectively.
\par
Let\vspace{-5pt}
\begin{equation}\Lambda^{\alpha}_{\mu\nu} :=
\Gamma^{\alpha}_{\mu\nu} - \Gamma^{\alpha}_{\nu\mu} =
\gamma^{\alpha}_{\mu\nu} - \gamma^{\alpha}_{\nu\mu}.\vspace{-5pt}
\end{equation} be the torsion tensor of the canonical connection
$\Gamma^{\alpha}_{\mu\nu}$ and\vspace{-5pt}
\begin{equation}\Omega^{\alpha}_{\mu\nu}: =
\gamma^{\alpha}_{\mu\nu} +
\gamma^{\alpha}_{\nu\mu}.\vspace{-5pt}\vspace{-5pt}
\end{equation}
Similarly, let\vspace{-5pt}
\begin{equation}T^{\alpha}_{\mu\nu}: = C^{\alpha}_{\mu\nu} - C^{\alpha}_{\nu\mu} =
G^{\alpha}_{\mu\nu} - G^{\alpha}_{\nu\mu}\vspace{-5pt}
\end{equation} be what we may call the
torsion tensor of $C^{\alpha}_{\mu\nu}$ and\vspace{-5pt}
\begin{equation}D^{\alpha}_{\mu\nu}: = G^{\alpha}_{\mu\nu} + G^{\alpha}_{\nu\mu}.\vspace{-5pt}\end{equation}
Now, if $\gamma_{\sigma\mu\nu}: =
g_{\epsilon\sigma}\gamma^{\epsilon}_{\mu\nu}$ and $G_{\sigma\mu\nu}:
= g_{\epsilon\sigma} G^{\epsilon}_{\mu\nu}$, then
$\gamma_{\sigma\mu\nu}$ and $G_{\sigma\mu\nu}$ are skew symmetric in
the first pair of indices. This, in turn, implies that\vspace{-5pt}
\begin{equation}\gamma^{\epsilon}_{\epsilon\nu} = G^{\epsilon}_{\epsilon\nu} = 0.\vspace{-5pt}\end{equation}
Hence, if\vspace{-5pt}
$$\beta_{\mu}: = \gamma^{\epsilon}_{\mu\epsilon}, \ \ B_{\mu}: = G^{\epsilon}_{\mu\epsilon},$$
then\vspace{-5pt}
\begin{equation}\Lambda^{\epsilon}_{\mu\epsilon} = \gamma^{\epsilon}_{\mu\epsilon} = \beta_{\mu}, \ \ \
T^{\epsilon}_{\mu\epsilon} = G^{\epsilon}_{\mu\epsilon} =
B_{\mu}.\end{equation} Finally, it can be shown, in analogy to the
classical AP-space \cite{HS}, that the contortion tensors
$\gamma_{\mu\nu\sigma}$ and $G_{\mu\nu\sigma}$ can be expressed in
terms of the torsion tensors in the form\vspace{-5pt}
\begin{equation}\gamma_{\mu\nu\sigma} = \frac{1}{2} (\Lambda_{\mu\nu\sigma} +
\Lambda_{\sigma\nu\mu} +
\Lambda_{\nu\sigma\mu})\vspace{-5pt}\end{equation}
\begin{equation}G_{\mu\nu\sigma} = \frac{1}{2} (T_{\mu\nu\sigma} +
T_{\sigma\nu\mu} + T_{\nu\sigma\mu}),\end{equation} where
$\Lambda_{\mu\nu\sigma}: =
g_{\epsilon\mu}\Lambda^{\epsilon}_{\nu\sigma}$ and
$T_{\mu\nu\sigma}: = g_{\epsilon\mu}T^{\epsilon}_{\nu\sigma}$. It is
clear by (3.11), (3.13), (3.17) and (3.18) that the torsion tensors
vanish if and only if the contortion tensors vanish.
\bigskip

The next  table gives a comparison between the fundamental geometric
objects in the classical AP-geometry  and the GAP-geometry. Similar
objects of the two spaces will be denoted by the same symbol. As
previously mentioned, \lq\lq $h$\rq\rq\, stands for \lq\lq
horizontal\rq\rq\, whereas \lq\lq $v$\rq\rq\,stands for \lq\lq
vertical\rq\rq. \vspace{5pt}
\begin{center} {\bf Table 1: Comparison between the classical AP-geometry and the
\\GAP-geometry}\\[0.3 cm]

\begin{tabular}
{|c|c|c|c|c|c|c|}\hline
&&\\
 \ \ &Classical AP-geometry&GAP-geometry
\\[0.2cm]\hline

&&\\
Building blocks&${\lambda}^{\alpha}(x)$&${\lambda}^{\alpha}(x, y)$
\\[0.2cm]\hline
&&\\
Metric&$g_{\mu\nu}(x) = \undersym{\lambda}{i}_{\mu}(x) \
\undersym{\lambda}{i}_{\nu}(x)$ &$g_{\mu\nu}(x, y) =
\undersym{\lambda}{i}_{\mu}(x, y) \ \undersym{\lambda}{i}_{\nu}(x,
y)$
\\[0.2cm]\hline
&&\\
Riemannian
connection&\footnotesize${\overcirc{\Gamma}^{\alpha}_{\mu\nu} =
\frac{1}{2}g^{\alpha\epsilon}\{\partial_{\mu}g_{\nu\epsilon} +
\partial_{\nu}g_{\mu\epsilon} + \partial_{\epsilon}g_{\mu\nu}\}}$&
\footnotesize${\overcirc{\Gamma}^{\alpha}_{\mu\nu} =
\frac{1}{2}g^{\alpha\epsilon}\{\delta_{\mu}g_{\nu\epsilon} +
\delta_{\nu}g_{\mu\epsilon} + \delta_{\epsilon}g_{\mu\nu}\}}$
\,($h$)
\\[0.01 cm]
&&\\
\ \ & \ \ & \footnotesize{$\overcirc{C}^{\alpha}_{\mu\nu} =
\frac{1}{2}g^{\alpha\epsilon}\{\dot{\partial}_{\mu}g_{\nu\epsilon} +
\dot{\partial}_{\nu}g_{\mu\epsilon} +
\dot{\partial}_{\epsilon}g_{\mu\nu}\}$ ($v$)}
\\[0.2cm]\hline
&&\\
Canonical connection&$\Gamma^{\alpha}_{\mu\nu} =
\undersym{\lambda}{i}^{\alpha} (\partial_{\nu} \
\undersym{\lambda}{i}_{\mu})$& $\Gamma^{\alpha}_{\mu\nu} =
\undersym{\lambda}{i}^{\alpha}(\delta_{\nu} \
\undersym{\lambda}{i}_{\mu})$ \ \,($h$-counterpart)
\\[0.01 cm]
&&\\
\ \ & \ \ & $C^{\alpha}_{\mu\nu} = \
\undersym{\lambda}{i}^{\alpha}(\dot{\partial_{\nu}} \
\undersym{\lambda}{i}_{\mu})$ \ ($v$-counterpart)
\\[0.2 cm]\hline
&&\\
AP-condition& ${\lambda}^{\alpha} \,\!_{|\mu} =
0$&${\lambda}^{\alpha}\, \!_{|\mu} = 0$ \ \,($h$-covariant
derivative)
\\[0.01 cm]
&&\\
\ \ & \ \ & ${\lambda}^{\alpha} \!_{||\mu} = \ 0$ \ ($v$-covariant
derivative)
\\[0.2 cm]\hline
&&\\
Torsion& $\Lambda^{\alpha}_{\mu\nu} = \Gamma^{\alpha}_{\mu\nu} -
\Gamma^{\alpha}_{\nu\mu}$&$\Lambda^{\alpha}_{\mu\nu} =
\Gamma^{\alpha}_{\mu\nu} - \Gamma^{\alpha}_{\nu\mu}$ \
($h$-counterpart)
\\[0.01 cm]
&&\\
\ \ & \ \ & $T^{\alpha}_{\mu\nu} = C^{\alpha}_{\mu\nu} -
C^{\alpha}_{\nu\mu}$ \ ($v$-counterpart)
\\[0.2 cm]\hline
&&\\
Contorsion& $\gamma^{\alpha}_{\mu\nu} = \Gamma^{\alpha}_{\mu\nu} - \
\overcirc{\Gamma}^{\alpha}_{\mu\nu}$& $\gamma^{\alpha}_{\mu\nu} =
\Gamma^{\alpha}_{\mu\nu} - \ \overcirc{\Gamma}^{\alpha}_{\nu\mu}$ \
\,\,($h$-counterpart)
\\[0.01 cm]
&&\\
\ \ & \ \ & $G^{\alpha}_{\mu\nu} = C^{\alpha}_{\mu\nu} - \
\overcirc{C}^{\alpha}_{\mu\nu}$ \ ($v$-counterpart)
\\[0.2 cm]\hline
&&\\
Basic vector& $\beta_{\mu} = \Lambda^{\alpha}_{\mu\alpha} =
\gamma^{\alpha}_{\mu\alpha}$& $\beta_{\mu} =
\Lambda^{\alpha}_{\mu\alpha} = \gamma^{\alpha}_{\mu\alpha}$
\,\,\,\,($h$-counterpart)
\\[0.01 cm]
&&\\
\ \ & \ \ & $B_{\mu} = T^{\alpha}_{\mu\alpha} =
G^{\alpha}_{\mu\alpha}$ \,\,($v$-counterpart)
\\[0.2 cm]\hline
\end{tabular}
\end{center}

%%%%%%%%%%%%%%%%%%%%%%%%%%%%%%%%%%%%%%%%%%%%% Section 4: Curvature tensors %%%%%%%%%%%%%%%%%%%%%%%%%%%%%%%%%%%%%%%%%%%%%%%%%%%%%

\newpage

\Section{Curvature tensors in Generalized AP-space}

Owing to the existence of two types of covariant derivatives with
respect to the canonical connection $D$, we have essentially three
commutation formulae and consequently three curvature tensors.
\begin{lem} Let $[\delta_{\sigma}, \delta_{\mu}] := \delta_{\sigma}\delta_{\mu} -
\delta_{\mu}\delta_{\sigma}$ and let $[\delta_{\sigma},
\dot{\partial}_{\mu}]$ be similarly defined. Then
\begin{equation}[\delta_{\sigma}, \delta_{\mu}] = R^{\epsilon}_{\sigma\mu} \ \dot{\partial}_{\epsilon},
\ \ [\delta_{\sigma}, \dot{\partial}_{\mu}] =
(\dot{\partial}_{\mu}N^{\epsilon}_{\sigma}) \
\dot{\partial}_{\epsilon},\end{equation} where
$R^{\alpha}_{\sigma\mu}: = \delta_{\mu}N^{\alpha}_{\sigma} -
\delta_{\sigma} N^{\alpha}_{\mu}$ is the curvature tensor of the
non-linear connection $N^{\alpha}_{\mu}$.
\end{lem}

\begin{thm} The three commutation formulae of \ $\undersym{\lambda}{i}^{\alpha}$
corresponding to the canonical connection $D = (N^{\alpha}_{\mu},
\Gamma^{\alpha}_{\mu\nu}, C^{\alpha}_{\mu\nu})$ are given by
\begin{description}
\item[(a)] ${\lambda}^{\alpha}\!\, _{|\mu\sigma} -
{\lambda}^{\alpha}\! \,_{|\sigma\mu} =
{\lambda}^{\epsilon}\,R^{\alpha}_{\epsilon\mu\sigma} +
{\lambda}^{\alpha}\!\, _{|\epsilon}\,\Lambda^{\epsilon}_{\sigma\mu}
+ {\lambda}^{\alpha}\!\, _{||\epsilon}\,R^{\epsilon}_{\sigma\mu}$
\item[(b)] ${\lambda}^{\alpha}\!\,_{||\mu\sigma} -
{\lambda}^{\alpha}\!\,_{||\sigma\mu} = {\lambda}^{\epsilon}
S^{\alpha}_{\epsilon\mu\sigma} + {\lambda}^{\alpha}
\!\,_{||\epsilon}\,T^{\epsilon}_{\sigma\mu}$
\item[(c)] ${\lambda}^{\alpha}\!\, _{||\mu|\sigma} -
{\lambda}^{\alpha}\!\,_{|\sigma||\mu} = {\lambda}^{\epsilon}
P^{\alpha}_{\epsilon\mu\sigma} +
{\lambda}^{\alpha}\!\,_{|\epsilon}\,C^{\epsilon}_{\sigma\mu} + \
{\lambda}^{\alpha}\!\,_{||\epsilon}\,P^{\epsilon}_{\sigma\mu},$
\end{description}
where
\\[ - 0.7 cm]
\begin{eqnarray*}
R^{\alpha}_{\nu\mu\sigma}: &=& (\delta_{\sigma}
\Gamma^{\alpha}_{\nu\mu} - \delta_{\mu}\Gamma^{\alpha}_{\nu\sigma})
 + (\Gamma^{\epsilon}_{\nu\mu}\Gamma^{\alpha}_{\epsilon\sigma} -
\Gamma^{\epsilon}_{\nu\sigma}\Gamma^{\alpha}_{\epsilon\mu}) +
L^{\alpha}_{\nu\mu\sigma}, \ \text{(h-curvature)}\\
S^{\alpha}_{\nu\mu\sigma}: &=& \dot{\partial}_{\sigma}
C^{\alpha}_{\nu\mu} -\dot{\partial}_{\mu} C^{\alpha}_{\nu\sigma} +
C^{\epsilon}_{\nu\mu}C^{\alpha}_{\epsilon\sigma} -
C^{\epsilon}_{\nu\sigma}C^{\alpha}_{\epsilon\mu}, \ \ \ \ \ \ \ \ \ \ \ \ \text{(v-curvature)}\\
P^{\alpha}_{\nu\mu\sigma}: &=& C^{\alpha}_{\nu\mu|\sigma} -
\dot{\partial}_{\mu} \Gamma^{\alpha}_{\nu\sigma} -
P^{\epsilon}_{\sigma\mu}
C^{\alpha}_{\nu\epsilon}, \ \ \ \ \ \ \ \ \ \ \ \ \ \ \ \ \ \ \ \ \ \ \ \ \ \text{(hv-curvature)}\\
\end{eqnarray*}
\\[ -1.2 cm] given that $\,\,\,L^{\alpha}_{\nu\mu\sigma}: =
C^{\alpha}_{\nu\epsilon}\,R^{\epsilon}_{\mu\sigma}\,\,\,$ and
$\,\,\,P^{\nu}_{\sigma\mu}: = \dot{\partial}_{\mu} N^{\nu}_{\sigma}
- \Gamma^{\nu}_{\mu\sigma}$.
\end{thm}

A direct consequence of the above commutation formulae, together
with the fact that \ ${\lambda}^{\alpha}\!\,_{|\mu} =
{\lambda}^{\alpha}\!\, _{||\mu} = 0,$ is the following
\begin{cor} The three curvature tensors $R^{\alpha}_{\nu\mu\sigma}$,
$S^{\alpha}_{\nu\mu\sigma}$ and $P^{\alpha}_{\nu\mu\sigma}$ of the
canonical connection $D = (N^{\alpha}_{\mu},
\Gamma^{\alpha}_{\mu\nu}, C^{\alpha}_{\mu\nu})$ vanish identically.
\end{cor}
It is to be noted that the above result is a natural generalization
of the corresponding result of the classical AP-geometry \cite{AMR}.

\bigskip

The Bianchi identities \cite{MM} for the canonical $d$-connection
$(N^{\alpha}_{\mu}, \Gamma^{\alpha}_{\mu\nu}, C^{\alpha}_{\mu\nu})$
gives

\begin{prop} The following identities hold
\begin{description}
\item[(a)] $\mathfrak{S}_{\nu, \mu, \sigma} \Lambda^{\alpha}_{\nu\mu|\sigma} =
\mathfrak{S}_{\nu, \mu, \sigma}
(\Lambda^{\alpha}_{\mu\epsilon}\Lambda^{\epsilon}_{\nu\sigma} +
L^{\alpha}_{\mu\nu\sigma})$
\item[(b)] $\mathfrak{S}_{\nu, \mu, \sigma} T^{\alpha}_{\nu\mu||\sigma} =
\mathfrak{S}_{\nu, \mu, \sigma}
(T^{\alpha}_{\mu\epsilon}T^{\epsilon}_{\nu\sigma}),$

where $\mathfrak{S}_{\nu, \mu, \sigma}$ denotes a cyclic permutation
on $\nu, \mu, \sigma$.
\end{description}

\end{prop}

\begin{cor} The following identities hold:
\begin{description}

\item[(a)] $\Lambda^{\epsilon}_{\mu\nu|\epsilon} = \beta_{\mu|\nu} -
\beta_{\nu|\mu} +
\beta_{\epsilon}\Lambda^{\epsilon}_{\mu\nu}+\mathfrak{S}_{\epsilon,
\nu,\mu} L^{\epsilon}_{\epsilon\nu\mu}.$

\item[(b)] $T^{\epsilon}_{\mu\nu||\epsilon} = B_{\mu||\nu} -
B_{\nu||\mu} + B_{\epsilon}T^{\epsilon}_{\mu\nu},$
\end{description}
\end{cor}
%\begin{proof}
\prof  Both identities follow by contracting the indices $\alpha$
and $\sigma$ in the identities (a) and (b) of Proposition 4.4,
taking into account that $\beta_{\mu} =
\Lambda^{\epsilon}_{\mu\epsilon}$, $B_{\mu} =
T^{\epsilon}_{\mu\epsilon}$ and $L^{\alpha}_{\mu\nu\sigma} = -
L^{\alpha}_{\mu\sigma\nu}.$ \ \ $\Box$
%\end{proof}

\bigskip

In addition to the Riemannian and the cannonical $d$-connections,
our space\linebreak admits at least two other natural
$d$-connections. In analogy to the classical AP-space, we define the
dual $d$-connection $\widetilde{D} = (N^{\alpha}_{\mu},
\widetilde{\Gamma}^{\alpha}_{\mu\nu},
\widetilde{C}^{\alpha}_{\mu\nu})$ by
\begin{equation}\widetilde{\Gamma}^{\alpha}_{\mu\nu} :=
 \Gamma^{\alpha}_{\nu\mu}, \ \
\widetilde{C}^{\alpha}_{\mu\nu} := C^{\alpha}_{\nu\mu}\end{equation}
and the symmetric $d-$connection $\widehat {D} = (N^{\alpha}_{\mu},
\widehat{\Gamma}^{\alpha}_{\mu\nu}, \widehat{C}^{\alpha}_{\mu\nu})$
by
\begin{equation}\widehat{\Gamma}^{\alpha}_{\mu\nu}: =
\frac{1}{2}(\Gamma^{\alpha}_{\mu\nu} + \Gamma^{\alpha}_{\nu\mu}), \
\ \widehat {C}^{\alpha}_{\mu\nu}: = \frac{1}{2}(C^{\alpha}_{\mu\nu}
+ C^{\alpha}_{\nu\mu}).\end{equation} Covariant differentiation with
respect to $\widetilde{\Gamma}^{\alpha}_{\mu\nu}$ and
$\widehat{\Gamma}^{\alpha}_{\mu\nu}$ will be denoted by
\lq\lq\,$\widetilde{|}\,"$ and \lq\lq\,$\widehat{|}\,"$
respectively.

\bigskip

Now, corresonding to each of the four $d$-connections there are
three curvature tensors. Therefore, we have a total of twelve
curvature tensors three of which, as already mentioned, vanish
identically. The vanishing of the curvature tensors of the canonical
$d$-connection allows us to express, in a relatively compact form,
six of the other curvature tensors (the $h$- and $v$-curvature
tensors) corresponding to the Riemannian, symmetric and the dual
$d$-connections. These curvature tensors are expressed in terms of
the torsion tensors $\Lambda^{\alpha}_{\mu\nu}$,
$T^{\alpha}_{\mu\nu}$ and their covariant derivatives with respect
to the canonical $d$-connection, together with the curvature
$R^{\alpha}_{\mu\nu}$ of the non-linear connection
$N^{\alpha}_{\mu}.$ The other three $hv$-curvature tensors are
calculated, though their expressions are more complicated. This is
to be expected since the expression obtained for the $hv$-curvature
tensor of the canonical $d$-connection lacks the symmetry properties
enjoyed by the $h$- and $v$-curvature tensors.

\begin{thm} The h-, v- and hv-curvature tensors of the dual d-connection
$\widetilde{D} = (N^{\alpha}_{\mu},
\widetilde{\Gamma}^{\alpha}_{\mu\nu},
\widetilde{C}^{\alpha}_{\mu\nu})$ can be expressed in the form:
\begin{description}
\item[(a)] $\widetilde{R}^{\alpha}_{\mu\sigma\nu} =
\Lambda^{\alpha}_{\sigma\nu|\mu} + C^{\alpha}_{\epsilon\mu}
R^{\epsilon}_{\sigma\nu} + L^{\alpha}_{\sigma\nu\mu} +
L^{\alpha}_{\nu\mu\sigma}.$
\item[(b)] $\widetilde{S}^{\alpha}_{\mu\sigma\nu} = T^{\alpha}_{\sigma\nu||\mu}.$
\item[(c)] $\widetilde{P}^{\alpha}_{\nu\mu\sigma} = T^{\alpha}_{\mu\nu|\sigma} -
\Lambda^{\alpha}_{\sigma\nu||\mu} + T^{\epsilon}_{\mu\nu}
\Lambda^{\alpha}_{\sigma\epsilon} - T^{\alpha}_{\mu\epsilon}
\Lambda^{\epsilon}_{\sigma\nu} -
\Lambda^{\alpha}_{\epsilon\nu}C^{\epsilon}_{\sigma\mu} -
P^{\epsilon}_{\sigma\mu} T^{\alpha}_{\epsilon\nu}.$
\end{description}
The corresponding curvature tensors of the symmetric d-connection
$\widehat {D} = (N^{\alpha}_{\mu},
\widehat{\Gamma}^{\alpha}_{\mu\nu}, \widehat C^{\alpha}_{\mu\nu})$
can be expressed in the form:
\begin{description}
\item[(d)] $\widehat R^{\alpha}_{\mu\sigma\nu} =
\frac{1}{2}(\Lambda^{\alpha}_{\mu\nu|\sigma} -
\Lambda^{\alpha}_{\mu\sigma|\nu}) +
\frac{1}{4}(\Lambda^{\epsilon}_{\mu\nu}\Lambda^{\alpha}_{\sigma\epsilon}
- \Lambda^{\epsilon}_{\mu\sigma}\Lambda^{\alpha}_{\nu\epsilon})
 + \frac{1}{2}(\Lambda^{\epsilon}_{\sigma\nu}\Lambda^{\alpha}_{\epsilon\mu})
+ \frac{1}{2} (T^{\alpha}_{\epsilon\mu}R^{\epsilon}_{\sigma\nu}).$
\item[(e)] $\widehat S^{\alpha}_{\mu\sigma\nu} = \frac{1}{2}(T^{\alpha}_{\mu\nu||\sigma} -
T^{\alpha}_{\mu\sigma||\nu}) +
\frac{1}{4}(T^{\epsilon}_{\mu\nu}T^{\alpha}_{\sigma\epsilon} -
T^{\epsilon}_{\mu\sigma}T^{\alpha}_{\nu\epsilon})
 + \frac{1}{2}(T^{\epsilon}_{\sigma\nu}T^{\alpha}_{\epsilon\mu}).$
\item[(f)] $\widehat{P}^{\alpha}_{\nu\mu\sigma} = \frac{1}{2} \ (\Lambda^{\alpha}_{\mu\nu|\sigma} - \Lambda^{\alpha}_{\sigma\nu||\mu}) +
\frac{1}{4} \ \Lambda^{\epsilon}_{\sigma\mu}
T^{\alpha}_{\epsilon\nu} - \frac{1}{2} \
\Lambda^{\alpha}_{\epsilon\nu} C^{\epsilon}_{\sigma\mu} +
\frac{1}{4} \ \mathfrak{S}_{\mu, \nu,
\sigma}\Lambda^{\epsilon}_{\mu\nu} \Lambda^{\alpha}_{\sigma\epsilon}
- \frac{1}{2} \ P^{\epsilon}_{\sigma\mu} T^{\alpha}_{\epsilon\nu}.$
\end{description}
The corresponding curvature tensors of the Riemannian d-connection \
$\overcirc{D} = (N^{\alpha}_{\mu}, \
\overcirc{\Gamma}^{\alpha}_{\mu\nu}, \
\overcirc{C}^{\alpha}_{\mu\nu})$ can be expressed in the form
\begin{description}
\item[(g)] $\overcirc{R}^{\alpha}_{\mu\sigma\nu} = \gamma^{\alpha}_{\mu\nu|\sigma} -
\gamma^{\alpha}_{\mu\sigma|\nu} + \gamma^{\epsilon}_{\mu\sigma}
\gamma^{\alpha}_{\epsilon\nu} - \gamma^{\epsilon}_{\mu\nu}
\gamma^{\alpha}_{\epsilon\sigma} + \gamma^{\alpha}_{\mu\epsilon}
\Lambda^{\epsilon}_{\nu\sigma} +  G^{\alpha}_{\mu\epsilon}
R^{\epsilon}_{\nu\sigma}.$
\item[(h)]  $\overcirc {S}^{\alpha}_{\mu\sigma\nu} = G^{\alpha}_{\mu\nu||\sigma} -
G^{\alpha}_{\mu\sigma||\nu} +
G^{\epsilon}_{\mu\sigma}G^{\alpha}_{\epsilon\nu} -
G^{\epsilon}_{\mu\nu} G^{\alpha}_{\epsilon\sigma} +
G^{\alpha}_{\mu\epsilon} T^{\epsilon}_{\nu\sigma}.$
\item[(i)] $\overcirc{P}^{\alpha}_{\nu\mu\sigma} = \dot{\partial}_{u}\gamma^{\alpha}_{\nu\sigma} -
G^{\alpha}_{\nu\mu|\sigma} + (G^{\epsilon}_{\nu\mu} -
C^{\epsilon}_{\nu\mu}) \gamma^{\alpha}_{\epsilon\sigma} -
(G^{\alpha}_{\epsilon\mu} -
C^{\alpha}_{\epsilon\mu})\gamma^{\epsilon}_{\nu\sigma} +
P^{\epsilon}_{\sigma\mu}G^{\alpha}_{\nu\epsilon}.$
\end{description}
\end{thm}

%\begin{proof}
 \prof
We prove (a) and (c) only. The proof of the other parts is similar.
\begin{description}
\item[(a)] We have
\begin{eqnarray*} \widetilde {R}^{\alpha}_{\mu\sigma\nu}&=&
\delta_{\nu}\widetilde{\Gamma}^{\alpha}_{\mu\sigma} -
\delta_{\sigma}{\widetilde{\Gamma}}^{\alpha}_{\mu\nu} +
{\widetilde{\Gamma}}^{\epsilon}_{\mu\sigma}{\widetilde{\Gamma}}^{\alpha}_{\epsilon\nu}
- \widetilde {\Gamma}^{\epsilon}_{\mu\nu}\widetilde
{\Gamma}^{\alpha}_{\epsilon\sigma}
 + \widetilde{C}^{\alpha}_{\mu\epsilon} R^{\epsilon}_{\sigma\nu}\\
&=& \delta_{\nu}{\Gamma}^{\alpha}_{\sigma\mu} - \delta_{\sigma}
{\Gamma}^{\alpha}_{\nu\mu} +
{\Gamma}^{\epsilon}_{\sigma\mu}{\Gamma}^{\alpha}_{\nu\epsilon} -
{\Gamma}^{\epsilon}_{\nu\mu}{\Gamma}^{\alpha}_{\sigma\epsilon}
+ C^{\alpha}_{\epsilon\mu} R^{\epsilon}_{\sigma\nu}\\
&=&\{\delta_{\nu}\Gamma^{\alpha}_{\sigma\mu} +
\Gamma^{\epsilon}_{\sigma\mu} (\Lambda^{\alpha}_{\nu\epsilon} +
\Gamma^{\alpha}_{\epsilon\nu})\} -
\{\delta_{\sigma}\Gamma^{\alpha}_{\nu\mu} +
\Gamma^{\epsilon}_{\nu\mu}(\Lambda^{\alpha}_{\sigma\epsilon} +
\Gamma^{\alpha}_{\epsilon\sigma})\}\\
&& + \ C^{\alpha}_{\epsilon\mu}R^{\epsilon}_{\sigma\nu}\\
&=& (\delta_{\nu}\Gamma^{\alpha}_{\sigma\mu} + \Gamma^{\epsilon}_
{\sigma\mu}\Gamma^{\alpha}_{\epsilon\nu}) -
(\delta_{\sigma}\Gamma^{\alpha}_{\nu\mu} +
\Gamma^{\epsilon}_{\nu\mu}\Gamma^{\alpha}_{\epsilon\sigma}) -
(\Gamma^{\epsilon}_ {\sigma\mu}\Lambda^{\alpha}_{\epsilon\nu} +
\Gamma^{\epsilon}_{\nu\mu}\Lambda
^{\alpha}_{\sigma\epsilon})\\
&& + \ C^{\alpha}_{\epsilon\mu}R^{\epsilon}_{\sigma\nu}\\
&=& (R^{\alpha}_{\sigma\mu\nu} -
C^{\alpha}_{\sigma\epsilon}R^{\epsilon}_{\mu\nu} +
\delta_{\mu}\Gamma^{\alpha}_{\sigma\nu} +
\Gamma^{\epsilon}_{\sigma\nu}\Gamma^{\alpha}_{\epsilon\mu}) -
(R^{\alpha}_{\nu\mu\sigma} - C^{\alpha}_{\nu\epsilon} R^{\epsilon}_{\mu\sigma}\\
&& + \ \delta_{\mu}\Gamma^{\alpha}_{\nu\sigma} + \
\Gamma^{\epsilon}_{\nu\sigma}\Gamma^{\alpha}_{\epsilon\mu}) -
(\Gamma^{\epsilon}_{\sigma\mu}\Lambda^{\alpha}_{\epsilon\nu} +
\Gamma^{\epsilon}_{\nu\mu}\Lambda^{\alpha}_{\sigma\epsilon}) +
C^{\alpha}_{\epsilon\mu}R^{\epsilon}_{\sigma\nu}.\\
&=& \delta_{\mu}\Lambda^{\alpha}_{\sigma\nu} +
\Gamma^{\alpha}_{\epsilon\mu}\Lambda^{\epsilon}_{\sigma\nu} -
\Gamma^{\epsilon}_{\sigma\mu}\Lambda^{\alpha}_{\epsilon\nu} -
\Gamma^{\epsilon}_{\nu\mu}\Lambda^{\alpha}_{\sigma\epsilon} +
C^{\alpha}_{\epsilon\mu} R^{\epsilon}_{\sigma\nu} +
C^{\alpha}_{\sigma\epsilon} R^{\epsilon}_{\nu\mu} +
C^{\alpha}_{\nu\epsilon}
R^{\epsilon}_{\mu\sigma}\\
&=& \Lambda^{\alpha}_{\sigma\nu|\mu} + C^{\alpha}_{\epsilon\mu}
R^{\epsilon}_{\sigma\nu} +
L^{\alpha}_{\sigma\nu\mu} + L^{\alpha}_{\nu\mu\sigma}.\\
\end{eqnarray*}
\\[- 2 cm]
\item[(c)] We have
\begin{eqnarray*} \widetilde{P}^{\alpha}_{\nu\mu\sigma} &=& C^{\alpha}_{\mu\nu\widetilde{|}\sigma} -
\dot{\partial}_{\mu}\Gamma^{\alpha}_{\sigma\nu} -
(\dot{\partial}_{\mu}
N^{\epsilon}_{\sigma} - \Gamma^{\epsilon}_{\sigma\mu})C^{\alpha}_{\epsilon\nu}\\
&=& C^{\alpha}_{\nu\mu|\sigma} +
(C^{\alpha}_{\mu\nu\widetilde{|}\sigma} -
C^{\alpha}_{\nu\mu|\sigma}) -
\dot{\partial}_{\mu}\Lambda^{\alpha}_{\sigma\nu} -
\dot{\partial}_{\mu} \Gamma^{\alpha}_{\nu\sigma} -
\dot{\partial}_{\mu}
N^{\epsilon}_{\sigma}(T^{\alpha}_{\epsilon\nu} + C^{\alpha}_{\nu\epsilon})\\
&& + \ (\Lambda^{\epsilon}_{\sigma\mu} +
\Gamma^{\epsilon}_{\mu\sigma})
(T^{\alpha}_{\epsilon\nu} + C^{\alpha}_{\nu\epsilon})\\
&=& P^{\alpha}_{\nu\mu\sigma} -
(\dot{\partial}_{\mu}N^{\epsilon}_{\sigma} -
\Gamma^{\epsilon}_{\mu\sigma}) T^{\alpha}_{\epsilon\nu} -
\dot{\partial}_{\mu}\Lambda^{\alpha}_{\sigma\nu} +
\Lambda^{\epsilon}_{\sigma\mu} C^{\alpha}_{\epsilon\nu} +
(C^{\alpha}_{\mu\nu\widetilde{|}\sigma} - C^{\alpha}_{\nu\mu|\sigma})\\
&=& (C^{\alpha}_{\mu\nu\widetilde{|}\sigma} -
C^{\alpha}_{\nu\mu|\sigma}) + \Lambda^{\epsilon}_{\sigma\mu}
C^{\alpha}_{\epsilon\nu} -
\dot{\partial}_{\mu}\Lambda^{\alpha}_{\sigma\nu} -
P^{\epsilon}_{\sigma\mu} T^{\alpha}_{\epsilon\nu}\\
&=& T^{\alpha}_{\mu\nu|\sigma} +
C^{\epsilon}_{\mu\nu}\Lambda^{\alpha}_ {\sigma\epsilon}  -
C^{\alpha}_{\mu\epsilon}\Lambda^{\epsilon}_{\sigma\nu} -
\dot{\partial}_{\mu}\Lambda^{\alpha}_{\sigma\nu} -
P^{\epsilon}_{\sigma\mu}
T^{\alpha}_{\epsilon\nu}\\
&=& T^{\alpha}_{\mu\nu|\sigma} -
\dot{\partial}_{\mu}\Lambda^{\alpha}_{\sigma\nu} +
(T^{\epsilon}_{\mu\nu} +
C^{\epsilon}_{\nu\mu})\Lambda^{\alpha}_{\sigma\epsilon} -
(T^{\alpha}_{\mu\epsilon} +
C^{\alpha}_{\epsilon\mu})\Lambda^{\epsilon}_{\sigma\nu} -
P^{\epsilon}_{\sigma\mu} T^{\alpha}_{\epsilon\nu}\\
&=& T^{\alpha}_{\mu\nu|\sigma} - \Lambda^{\alpha}_{\sigma\nu||\mu} +
T^{\epsilon}_{\mu\nu}\Lambda^{\alpha}_{\sigma\epsilon} -
T^{\alpha}_{\mu\epsilon} \Lambda^{\epsilon}_{\sigma\nu} -
\Lambda^{\alpha}_{\epsilon\nu} C^{\epsilon}_{\sigma\mu} -
P^{\epsilon}_{\sigma\mu} T^{\alpha}_{\epsilon\nu}. \ \ \ \Box
\end{eqnarray*}
\end{description}
%\end{proof}

%%%%%%%%%%%%%%%%%%%%%%%%%%%%%%%%%%%%%%%%%%%%%%%%% Section 5: Second rank tensors %%%%%%%%%%%%%%%%%%%%%%%%%55%%%%%%%%%%%%%%%%%%%%%
\vspace{-1 cm} \Section{Fundamental second rank tensors}

Due to the importance of second order symmetric and skew-symmetric
tensors in physical applications, we here list such tensors in Table
2 below. We regard these tensors as {\bf fundamental} since their
counterparts in the classical AP-context {\it play a key role in
physical applications.} Moreover, in the AP-geometry, most second
rank tensors which have physical significance can be expressed as a
linear combination of these fundamental tensors. The Table is
constructed as similar as possible to that given by Mikhail (cf.
\cite{FI}, Table 2), to facilitate comparison with the case of the
classical AP-geometry which has many physical applications \cite{b}.
Corresponding \lq\lq horizontal" and \lq\lq vertical\rq\rq\, tensors
are denoted by the same symbol with the\linebreak \lq\lq
vertical\rq\rq\, tensors barred. It is to be noted that all \lq\lq
vertical \rq\rq \ tensors have no counterpart in the classical
AP-context.

\begin{center}\bf{Table 2: Summary of the fundamental symmetric and skew-symmetric
second rank tensors}

\bigskip

\begin{tabular}
{|c|c|c|c|}\hline
\multicolumn{2}{|c|}{\hbox{ }} &\multicolumn{2}{c|}{\hbox{ }}\\
\multicolumn{2}{|c|}{Horizontal}&\multicolumn{2}{c|}{Vertical}
\\[0.4 cm]\hline
&&&\\
Skew-Symmetric&Symmetric&Skew-Symmetric&Symmetric
\\[0.4 cm]\hline
&&&\\
$\xi_{\mu\nu}: = \gamma_{\mu\nu}  \!^{\alpha} \!\,_{|\alpha}$&$ \ $&
$\o {\xi}_{\mu\nu}: = G_{\mu\nu}  \!^{\alpha} \!\,_{|\alpha}$&$ \ $
\\[0.4 cm]\hline
&&&\\
$\gamma_{\mu\nu}: = \beta_{\alpha}\gamma_{\mu\nu} \!^{\alpha}$& $ \
$&$\o {\gamma}_{\mu\nu} : = B_{\alpha}G_{\mu\nu} \!^{\alpha}$& $ \ $
\\[0.4 cm]\hline
&&&\\
$\eta_{\mu\nu} := \beta_{\epsilon}\,\Lambda^{\epsilon}_{\mu\nu}$&
$\phi_{\mu\nu} := \beta_{\epsilon}\,\Omega^{\epsilon}_{\mu\nu}$ &$\o
{\eta}_{\mu\nu} := B_{\epsilon}\,T^{\epsilon}_{\mu\nu}$& $\o
{\phi}_{\mu\nu} := B_{\epsilon}\,D^{\epsilon}_{\mu\nu}$
\\[0.4 cm]\hline
&&&\\
$\chi_{\mu\nu} := \Lambda^{\alpha}_{\mu\nu|\alpha}$& $\psi_{\mu\nu}
:= \Omega^{\epsilon}_{\mu\nu|\epsilon}$& $\o {\chi}_{\mu\nu} :=
T^{\alpha}_{\mu\nu||\alpha}$& $\o {\psi}_{\mu\nu} :=
D^{\alpha}_{\mu\nu||\alpha}$
\\[0.4 cm]\hline
&&&\\
$\epsilon_{\mu\nu} := \frac{1}{2}(\beta_{\mu|\nu} -
\beta_{\nu|\mu})$& $\theta_{\mu\nu} :=  \frac{1}{2}(\beta_{\mu|\nu}
+ \beta_{\nu|\mu})$& $\o {\epsilon}_{\mu\nu} :=
\frac{1}{2}(B_{\mu||\nu} - B_{\nu||\mu})$& $\o {\theta}_{\mu\nu} :=
\frac{1}{2}(B_{\mu||\nu} + B_{\nu||\mu})$
\\[0.4 cm]\hline
&&&\\
\footnotesize {$k_{\mu\nu} :=
\gamma^{\epsilon}_{\alpha\mu}\gamma^{\alpha}_{\nu\epsilon}
 - \gamma^{\epsilon}_{\mu\alpha}\gamma^{\alpha}_{\epsilon\nu}$}&
\footnotesize{$h_{\mu\nu}: =
\gamma^{\epsilon}_{\alpha\mu}\gamma^{\alpha}_{\nu\epsilon} +
\gamma^{\epsilon}_{\mu\alpha}\gamma^{\alpha}_{\epsilon\nu}$}&

\footnotesize{$\o {k}_{\mu\nu}: =
G^{\epsilon}_{\alpha\mu}G^{\alpha}_{\nu\epsilon} -
G^{\epsilon}_{\mu\alpha}G^{\alpha}_{\epsilon\nu}$}&
\footnotesize{$\o {h}_{\mu\nu} :=
G^{\epsilon}_{\alpha\mu}G^{\alpha}_{\nu\epsilon} +
G^{\epsilon}_{\mu\alpha}G^{\alpha}_{\epsilon\nu}$}
\\[0.4 cm]\hline
&&&\\
$ \ $&$\sigma_{\mu\nu} :=
\gamma^{\epsilon}_{\alpha\mu}\gamma^{\alpha}_{\epsilon\nu}$& $ \ $&
$\o {\sigma}_{\mu\nu}: =
G^{\epsilon}_{\alpha\mu}G^{\alpha}_{\epsilon\nu}$

\\[0.4 cm]\hline
&&&\\
$ \ $&$\omega_{\mu\nu} :=
\gamma^{\epsilon}_{\mu\alpha}\gamma^{\alpha}_{\nu\epsilon}$& $  \ $&
$\o {\omega}_{\mu\nu} :=
G^{\epsilon}_{\mu\alpha}G^{\alpha}_{\nu\epsilon}$
\\[0.4 cm]\hline
&&&\\
$ \ $&$\alpha_{\mu\nu} := \beta_{\mu}\beta_{\nu}$& $ \ $& $\o
{\alpha}_{\mu\nu} := B_{\mu}B_{\nu}$
\\[0.4 cm]\hline
\end{tabular}
\end{center}

Due to the metricity condition in Theorem 3.3, one can use the
metric tensor $g_{\mu\nu}$ and its inverse $g^{\mu\nu}$ to perform
the operations of lowering and raising tensor indices under the $h$-
and $v$- covariant derivatives relative to the canonical
$d$-connection.

Thus, contraction with the metric tensor of the above fundamental
tensors gives the following table of scalars:
\begin{center} {\bf Table 3: Summary of the fundamental scalars}\\[0.3cm]

\begin{tabular}
{|c|c|c|c|c|c|c|}\hline
&&&&\\
{\bf Horizontal}& \footnotesize${\alpha := \beta_{\mu}\beta^{\mu}}$&
\footnotesize${\theta := \beta^{\mu}
\,\!_{|\mu}}$&\footnotesize${\phi :=
\beta_{\epsilon}\,\Omega^{\epsilon\mu}
\,\!_{\mu}}$&\footnotesize${\psi := \Omega^{\alpha\mu}
\,\!_{\mu|\alpha}}$
\\[0.2cm]\cline{2-5}
 &&&&\\
 \ \  &\footnotesize{$\omega := \gamma^{\epsilon\mu} \!_{\alpha}
\,\gamma^{\alpha}_{\mu\epsilon}$}&\footnotesize${\sigma: =
\gamma^{\epsilon}
\!_{\alpha}\,\!^{\mu}\,\gamma^{\alpha}_{\epsilon\mu}}$&
\footnotesize{$h := 2\gamma^{\alpha\mu}
\!_{\epsilon}\,\gamma^{\epsilon}_{\alpha\mu}$}
 &\ \  \ \  \ \ \ \\[0.2cm]\hline
&&&&\\

{\bf Vertical}& \footnotesize${\o {\alpha} := B_{\mu}B^{\mu}}$&
\footnotesize${\o {\theta} := B^{\mu}
\,\!_{|\mu}}$&\footnotesize${\o {\phi} :=
B_{\epsilon}\,D^{\epsilon\mu} \!_{\mu}}$&\footnotesize${\o {\psi} :=
D^{\alpha\mu} \,\!_{\mu|\alpha}}$
\\[0.2cm]\cline{2-5}
 &&&&\\
 \ \ &\footnotesize{$\o {\omega} := G^{\epsilon\mu} \!_{\alpha}
\,G^{\alpha}_{\mu\epsilon}$}&\footnotesize${\o {\sigma} :=
G^{\epsilon} \!_{\alpha} \!^{\mu}\,G^{\alpha}_{\epsilon\mu}}$&
\footnotesize{$\o {h} := 2G^{\alpha\mu}
\!_{\epsilon}\,G^{\epsilon}_{\alpha\mu}$}
& \ \ \ \ \ \ \ \\[0.2cm]\hline
\end{tabular}
\end{center}

\bigskip

In physical applications, second order symmetric tensors of zero
trace have special importance. For example, in the case of
electromagnetism, the tensor characterizing the electro-magnetic
energy is a second order symmetric tensor having zero trace. So it
is of interest to search for such tensors. The Table below gives
some of the second rank tensors of zero trace.
%(As easily checked,
%we have $\phi + 2\alpha = \psi + 2\theta = h + 2\omega =
%\frac{1}{2}(\phi - \psi) + \theta - \alpha - 2\beta^{\alpha}
%\,_{\widetilde{|}\alpha} = 0\,$  and similarly for the vertical
%counterparts).
\begin{center}{\bf{Table 4: Summary of the fundamental tensors of zero trace}}

\bigskip

%\begin{center}{\bf{Table 2 (b)}}
%\end{center}
\begin{tabular}
{|c|c|}\hline
&\\
\bf{Horizontal}&\bf{Vertical}
\\[0.4 cm]\hline
&\\
$\phi_{\mu\nu} + 2\alpha_{\mu\nu}$&$\o {\phi}_{\mu\nu} +
2\bar{\alpha}_{\mu\nu}$
\\[0.4 cm]\hline
&\\
$\psi_{\mu\nu} + 2\theta_{\mu\nu}$&$\o {\psi}_{\mu\nu} + 2\o
{\theta}_{\mu\nu}$
\\[0.4 cm]\hline
&\\
$h_{\mu\nu} + 2\omega_{\mu\nu}$&$\o {h}_{\mu\nu} + 2\o
{\omega}_{\mu\nu}$
\\[0.4 cm]\hline
&\\
\footnotesize{$\frac{1}{2}(\phi_{\mu\nu} - \psi_{\mu\nu}) +
\theta_{\mu\nu} - \alpha_{\mu\nu} -
\frac{1}{2}g_{\mu\nu}\beta^{\alpha}
\,_{\widetilde{|}\alpha}$}&\footnotesize{$\frac{1}{2} (\o
{\phi}_{\mu\nu} - \o {\psi}_{\mu\nu}) + \o {\theta}_{\mu\nu} - \o
{\alpha}_{\mu\nu} - \frac{1}{2}g_{\mu\nu}B^{\alpha}
\,_{\widetilde{||}\alpha}$}
\\[0.4 cm]\hline
\end{tabular}
\end{center}
%%%%%%%%%%%%%%%%%%%%%%%%%%%%%%%%%%%%%%%%%
\bigskip
\par
We now consider some useful second rank tensors which are not
expressible in terms of the fundamental tensors appearing in Table
2. Unlike the tensors of Table 2, some of the tensors to be defined
below have no horizontal and vertical counterparts. To this end,
let\vspace{-5pt}
$$L_{\mu\nu} := L^{\alpha}_{\alpha\mu\nu} =
C^{\alpha}_{\alpha\epsilon}R^{\epsilon}_{\mu\nu}, \ \  M_{\mu\nu}: =
L^{\alpha}_{\mu\alpha\nu} =
C^{\alpha}_{\mu\epsilon}\,R^{\epsilon}_{\alpha\nu}, \ \ N_{\mu\nu}:
=C^{\alpha}_{\epsilon\mu}\, R^{\epsilon}_{\alpha\nu}, \ \
F_{\mu\nu}: = \
\overcirc{C}^{\alpha}_{\epsilon\mu}\,R^{\epsilon}_{\alpha\nu}.\vspace{-5pt}$$
Then, clearly\vspace{-5pt}
$$T_{\mu\nu}: = M_{\mu\nu} - N_{\mu\nu} = T^{\alpha}_{\mu\epsilon}
\,R^{\epsilon}_{\alpha\nu}, \ \ G_{\mu\nu}: = M_{\mu\nu} -
F_{\mu\nu} = G^{\alpha}_{\mu\epsilon}\, R^{\epsilon}_{\alpha\nu}, \
\ G_{\mu\nu} - T_{\mu\nu} =
G^{\alpha}_{\epsilon\mu}\,R^{\epsilon}_{\alpha\nu}.\vspace{-5pt}$$
%Moreover, if $\ T :=g^{\mu\nu} T_{\mu\nu}\ $ and $\ G: = g^{\mu\nu}
%G_{\mu\nu}\ $, then \vspace{-5pt}
%$$T = T^{\alpha\mu} \!_{\epsilon}\,R^{\epsilon}_{\alpha\mu}, \ \ \
%G = G^{\alpha\mu} \!_{\epsilon}\,R^{\epsilon}_{\alpha\mu}, \ \ \ G -
%T = G^{\alpha} \!_{\epsilon} \ \!
%\!^{\mu}\,R^{\epsilon}_{\alpha\mu}.\vspace{-5pt}$$
%Finally as easily checked,
%$$\Lambda^{\alpha}_{\mu\epsilon}\,\Lambda^{\epsilon}_{\alpha\nu} =
%h_{\mu\nu} - \sigma_{\mu\nu} - \omega_{\mu\nu}, \ \qquad
%T^{\alpha}_{\mu\epsilon}\,T^{\epsilon}_{\alpha\nu} = \bar{h}_{\mu\nu}
%- \bar{\sigma}_{\mu\nu} - \bar{\omega}_{\mu\nu},$$ so that
%$$\Lambda^{\alpha\mu} \!_{\epsilon}\,\Lambda^{\epsilon}_{\alpha\mu} =
%- (3\omega + \sigma), \ \qquad T^{\alpha\mu}\!_{\epsilon}\,T^{\epsilon}_{\alpha\mu} = - (3\bar{\omega} +
% \bar{\sigma}).$$
Finally, let $\,T :=g^{\mu\nu} T_{\mu\nu}\,$ and $\,G: =
g^{\mu\nu}G_{\mu\nu}$. By the above, we have the following:\\
{Symmetric second rank tensors:} $M_{(\mu\nu)}$, $\ N_{(\mu\nu)}$,
$\ F_{(\mu\nu)}$.\\ { Skew-symmetric second rank tensors:}
$M_{[\mu\nu]}$, $\ N_{[\mu\nu]}$, $\ F_{[\mu\nu]}$, $\ L_{\mu\nu}$.

%These tensors are {\it{not fundamental}} in the sense described
%above. Their significance lie only in the fact that they appear in
%the expressions obtained for the contracted curvature tensors. We
%singled them out just to write the contracted curvatures in a
%relatively compact form.

%%%%%%%%%%%%%%%%%%%%%%%%%%%%%%%%%%%%%%%%% Section 6: Contracted curv. tensors %%%%%%%%%%%%%%%%%%%%%%%%%%%%%%%%%%%%%%%%%%%%%%%%

\Section{Contracted curvatures and curvature scalars}

It may be convenient, for physical reasons, to consider second rank
tensors derived from the curvature tensors by contractions. It is
also of interest to reduce the number of these tensors to a minimum
which is fundamental (cf. Propositions 6.1 and 6.2).
\par
Contracting the indices $\alpha$ and $\mu$ in the expressions
obtained for the $h$- and $v$-curvature tensors in Theorem 4.6,
taking into account Corollary 4.5, we obtain
\begin{prop} Let $\widetilde{\cal R}_{\sigma\nu} :=
\widetilde{R}^{\alpha}_{\alpha\sigma\nu}$, $\widehat{\cal
R}_{\sigma\nu} := \widehat{R}^{\alpha}_{\alpha\sigma\nu}$ and \
$\overcirc{\cal R}_{\sigma\nu}: = \
\overcirc{R}^{\alpha}_{\alpha\sigma\nu}$ with similar expressions
for $\widetilde{\cal S}_{\sigma\nu}$, $\widehat{\cal S}_{\sigma\nu}$
and \ $\overcirc{\cal S}_{\sigma\nu}$. Then, we have
\begin{description}
\item[(a)] $\widetilde{\cal {R}}_{\sigma\nu} =
\beta_{\sigma|\nu} - \beta_{\nu|\sigma} + \beta_{\epsilon}
\Lambda^{\epsilon}_{\sigma\nu} +
B_{\epsilon}R^{\epsilon}_{\sigma\nu},$
\item[(b)] $\widetilde{\cal {S}}_{\sigma\nu} =
B_{\sigma||\nu} - B_{\nu||\sigma} +
B_{\epsilon}T^{\epsilon}_{\sigma\nu},$
\item[(c)] $\widehat{\cal {R}}_{\sigma\nu} =
\frac{1}{2}\widetilde{\cal {R}}_{\sigma\nu},$
\item[(d)] $\widehat{\cal {S}}_{\sigma\nu} =
\frac{1}{2}\widetilde{\cal {S}}_{\sigma\nu},$
\item [(e)] $\overcirc{\cal R}_{\sigma\nu} = \ \overcirc{\cal S}_{\sigma\nu} = 0.$
\end{description}
\end{prop}

\begin{prop} Let $\widetilde{R}_{\mu\sigma} := \widetilde{R}^{\alpha}_{\mu\sigma\alpha}$,
$\widehat{R}_{\mu\sigma} := \widehat{R}^{\alpha}_{\mu\sigma\alpha}$
and \ $\overcirc{R}_{\mu\sigma} := \
\overcirc{R}^{\alpha}_{\mu\sigma\alpha}$ with similar expressions
for $\widetilde{S}_{\mu\sigma}$, $\widehat{S}_{\mu\sigma}$ and \
$\overcirc{S}_{\mu\sigma}$. Then, we have
\begin{description}
\item[(a)] $\widetilde{R}_{\mu\sigma} = \beta_{\sigma|\mu} + C^{\alpha}_{\epsilon\mu}
R^{\epsilon}_{\sigma\alpha} + L^{\alpha}_{\sigma\alpha\mu} +
L^{\alpha}_{\alpha\mu\sigma},$

\item[(b)] $\widetilde{S}_{\mu\sigma} = B_{\sigma||\mu},$

\item[(c)] $\widehat{R}_{\mu\sigma} = \frac{1}{2} \widetilde{R}_{\mu\sigma} + \frac{1}{4}\{
\beta_{\epsilon}\Lambda^{\epsilon}_{\sigma\mu} +
\Lambda^{\epsilon}_{\alpha\sigma}\Lambda^{\alpha}_{\mu\epsilon}\},$

\item[(d)] $\widehat{S}_{\mu\sigma} = \frac{1}{2} \widetilde{S}_{\mu\sigma} + \frac{1}{4}\{
B_{\epsilon}T^{\epsilon}_{\sigma\mu} +
T^{\epsilon}_{\alpha\sigma}T^{\alpha}_{\mu\epsilon}\},$

\item[(e)] $\overcirc{R}_{\mu\sigma} = \beta_{\mu|\sigma} - \gamma^{\alpha}_{\mu\sigma|\alpha} + \beta_{\epsilon}
\gamma^{\epsilon}_{\mu\sigma} - \gamma^{\alpha}_{\mu\epsilon}
\gamma^{\epsilon}_{\sigma\alpha} +  G^{\alpha}_{\mu\epsilon}
R^{\epsilon}_{\alpha\sigma},$
\item[(f)] $\overcirc{S}_{\mu\sigma}: = \ \overcirc{S}^{\alpha}_{\mu\sigma\alpha}
= B_{\mu||\sigma} - G^{\alpha}_{\mu\sigma||\alpha} + B_{\epsilon}
G^{\epsilon}_{\mu\sigma} - G^{\alpha}_{\mu\epsilon}
G^{\epsilon}_{\sigma\alpha}.$
\end{description}
\end{prop}

%%%%%%%%%%%%%%%%%%%%%%%%%
\begin{prop} The following holds.
\begin{description}
\item[(a)] $\widetilde{R}_{[\mu\sigma]} = \frac{1}{2}\{\beta_{\sigma|\mu} - \beta_{\mu|\sigma}\} +
C^{\epsilon}_{\epsilon\alpha}R^{\alpha}_{\mu\sigma} +
C^{\epsilon}_{(\alpha\sigma)} R^{\alpha}_{\epsilon\mu} -
C^{\epsilon}_{(\alpha\mu)} R^{\alpha}_{\epsilon\sigma},$
\item[(b)] $\widetilde{R}_{(\mu\sigma)} = \frac{1}{2}\{\beta_{\sigma|\mu} + \beta_{\mu|\sigma} +
T^{\epsilon}_{\alpha\mu}R^{\alpha}_{\sigma\epsilon} +
T^{\epsilon}_{\alpha\sigma} R^{\alpha}_{\mu\epsilon}\},$
\item[(c)] $\widetilde{S}_{[\mu\sigma]} = \frac{1}{2}\{B_{\sigma||\mu} - B_{\mu||\sigma}\},$

\item[(d)] $\widetilde{S}_{(\mu\sigma)} = \frac{1}{2}\{B_{\sigma||\mu} + B_{\mu||\sigma}\},$

\item[(e)] $\widehat{R}_{[\mu\sigma]} = \frac{1}{2} \widetilde{R}_{[\mu\sigma]} + \frac{1}{4}
\beta_{\epsilon}\,\Lambda^{\epsilon}_{\sigma\mu},$

\item[(f)] $\widehat{R}_{(\mu\sigma)} = \frac{1}{2} \widetilde{R}_{(\mu\sigma)} + \frac{1}{4}
\Lambda^{\epsilon}_{\alpha\sigma}\,\Lambda^{\alpha}_{\mu\epsilon},$

\item[(g)] $\widehat{S}_{[\mu\sigma]} = \frac{1}{2} \widetilde{S}_{[\mu\sigma]} + \frac{1}{4}
B_{\epsilon}\,T^{\epsilon}_{\sigma\mu},$

\item[(h)] $\widehat{S}_{(\mu\sigma)} = \frac{1}{2} \widetilde{S}_{(\mu\sigma)} + \frac{1}{4}
T^{\epsilon}_{\alpha\sigma}\,T^{\alpha}_{\mu\epsilon},$

\item[(i)] $\overcirc{R}_{[\mu\sigma]} = \frac{1}{2}\{L^{\alpha}_{\alpha\mu\sigma} +
\ \overcirc{C}^{\alpha}_{\sigma\epsilon}\,R^{\epsilon}_{\alpha\mu} -
\
\overcirc{C}^{\alpha}_{\mu\epsilon}\,R^{\epsilon}_{\alpha\sigma}\},$

\item[(j)] $\overcirc{R}_{(\mu\sigma)} = \frac{1}{2}\{(\beta_{\mu|\sigma} + \beta_{\sigma|\mu}) -
\Omega^{\alpha}_{\mu\sigma|\alpha} +
\beta_{\epsilon}\,\Omega^{\epsilon}_{\mu\sigma}\} -
\gamma^{\alpha}_{\mu\epsilon}\,\gamma^{\epsilon}_{\sigma\alpha} +
\frac{1}{2}\{G^{\alpha}_{\mu\epsilon}\,R^{\epsilon}_{\alpha\sigma} +
G^{\alpha}_{\sigma\epsilon}\, R^{\epsilon}_{\alpha\mu}\},$

\item[(k)] $\overcirc{S}_{[\mu\sigma]} = 0,$

\item[(l)] $\overcirc{S}_{(\mu\sigma)} = \frac{1}{2}\{(B_{\mu||\sigma} + B_{\sigma||\mu}) -
D^{\alpha}_{\mu\sigma||\alpha} +
B_{\epsilon}\,D^{\epsilon}_{\mu\sigma}\} -
G^{\alpha}_{\mu\epsilon}\,G^{\epsilon}_{\sigma\alpha}.$
\end{description}
%Consequently, \ $\overcirc{S}_{\mu\sigma}$ is symmetric.
\end{prop}
%%%%
%Taking the trace of the above tensors, we get
\begin{cor} The following holds:
\begin{description}
\item[(a)] $\widetilde{R}^{\sigma}_{\sigma} := g^{\mu\sigma}\widetilde{R}_{\mu\sigma} =
\beta^{\sigma} \!\,_{|\sigma} + T^{\epsilon\sigma}
\!\!\,_{\alpha}\,R^{\alpha}_{\epsilon\sigma},$
\item[(b)] $\widetilde{S}^{\sigma}_{\sigma} := g^{\mu\sigma}\widetilde{S}_{\mu\sigma} = B^{\sigma} \!_{||\sigma},$
\item[(c)] $\widehat{R}^{\sigma}_{\sigma} := g^{\mu\sigma}\widehat{R}_{\mu\sigma} =
\frac{1}{2}\{\beta^{\sigma} \!\,_{|\sigma} + T^{\epsilon\sigma}
\!\!\,_{\alpha}\,R^{\alpha}_ {\epsilon\sigma}\} +
\frac{1}{4}\Lambda^{\epsilon\sigma} \!\!\,_{\alpha}\,
\Lambda^{\alpha}_{\epsilon\sigma},$
\item[(d)] $\widehat{S}^{\sigma}_{\sigma} := g^{\mu\sigma}\widehat{S}_{\mu\sigma} =
\frac{1}{2}B^{\sigma} \!\,_{||\sigma} +
\frac{1}{4}T^{\epsilon\sigma} \!\!\,_{\alpha}\,T^{\alpha}_
{\epsilon\sigma},$
\item[(e)] $\overcirc{R}^{\sigma} \!_{\sigma} :=
g^{\mu\sigma} \ \overcirc{R}_{\mu\sigma} = \beta^{\sigma}
\!_{|\sigma} - \frac{1}{2}\Omega^{\alpha\sigma} \!_{\sigma|\alpha} +
\frac{1}{2}\beta_{\alpha}\, \Omega^{\alpha\sigma} \!_{\sigma} -
\gamma^{\alpha\sigma} \!_{\epsilon}
\,\gamma^{\epsilon}_{\sigma\alpha} +  G^{\alpha\sigma}
\!_{\epsilon}\, R^{\epsilon}_{\alpha\sigma},$
\item[(f)] $\overcirc{S}^{\sigma} \!_{\sigma} := g^{\mu\sigma} \ \overcirc{S}_{\mu\sigma} =
B^{\sigma} \!_{||\sigma} - \frac{1}{2}D^{\alpha\sigma}
\!_{\sigma||\alpha} + \frac{1}{2}B_{\alpha} \,D^{\alpha\sigma}
\!_{\sigma} - G^{\alpha\sigma} \!_{\epsilon}\,
G^{\epsilon}_{\sigma\alpha}.$
\end{description}
\end{cor}

We now apply a different method for calculating both \
$\overcirc{R}_{\mu\sigma}$ and \ $\overcirc{S}_{\mu\sigma}$, now
expressed in terms of the covariant derivative of the contorsion
tensors with respect to the {\it Riemannian} $d$-connection. Then we
obtain

\begin{prop} The \lq\lq Ricci" tensors \ $\overcirc{R}_{\mu\sigma}$ and \ $\overcirc{S}_{\mu\sigma}$
can be expressed in the form
\begin{description}
\item[(a)] $\ \overcirc{R}_{\mu\sigma} = \beta_{\mu{o\atop |}\sigma} -
\gamma^{\alpha}_{\mu\sigma{o\atop |}\alpha} -
\beta_{\epsilon}\gamma^{\epsilon}_{\mu\sigma} +
\gamma^{\epsilon}_{\mu\alpha} \gamma^{\alpha}_{\epsilon\sigma} +
G^{\alpha}_{\mu\epsilon}R^{\epsilon}_{\alpha\sigma}.$
\item[(b)] $ \overcirc{S}_{\mu\sigma} = B_{\mu{o\atop ||}\sigma} -
G^{\alpha}_{\mu\sigma{o\atop ||}\alpha} -
B_{\epsilon}G^{\epsilon}_{\mu\sigma} + G^{\epsilon}_{\mu\alpha}
G^{\alpha}_{\epsilon\sigma}.$
\end{description}
\end{prop}

%\begin{proof}
\prof We prove \textbf{(a)} only; the proof of \textbf{(b)} is
similar. \\We have
\begin{eqnarray*} 0 = R^{\alpha}_{\mu\sigma\alpha} &=& (\delta_{\alpha}
\Gamma^{\alpha}_{\mu\sigma} -
\delta_{\sigma}\Gamma^{\alpha}_{\mu\alpha})
 + (\Gamma^{\epsilon}_{\mu\sigma}\Gamma^{\alpha}_{\epsilon\alpha} -
\Gamma^{\epsilon}_{\mu\alpha}\Gamma^{\alpha}_{\epsilon\sigma}) +
R^{\epsilon}_{\sigma\alpha}C^{\alpha}_{\mu\epsilon}\\
&=& \delta_{\alpha}( \ \overcirc{\Gamma}^{\alpha}_{\mu\sigma} +
\gamma^{\alpha}_{\mu\sigma}) - \delta_{\sigma}( \
\overcirc{\Gamma}^{\alpha}_{\mu\alpha} +
\gamma^{\alpha}_{\mu\alpha}) + ( \
\overcirc{\Gamma}^{\epsilon}_{\mu\sigma} +
\gamma^{\epsilon}_{\mu\sigma})
( \ \overcirc{\Gamma}^{\alpha}_{\epsilon\alpha} + \gamma^{\alpha}_{\epsilon\alpha})\\
&&- \ ( \ \overcirc{\Gamma}^{\epsilon}_{\mu\alpha} +
\gamma^{\epsilon}_{\mu\alpha}) ( \
\overcirc{\Gamma}^{\alpha}_{\epsilon\sigma} +
\gamma^{\alpha}_{\epsilon\sigma}) +
R^{\epsilon}_{\sigma\alpha}C^{\alpha}_{\mu\epsilon}\\
&=& \overcirc{R}_{\mu\sigma} -
(\delta_{\sigma}\gamma^{\alpha}_{\mu\alpha} -
\gamma^{\alpha}_{\epsilon\alpha} \
\overcirc{\Gamma}^{\epsilon}_{\mu\sigma}) +
(\delta_{\alpha}\gamma^{\alpha}_{\mu\sigma} +
\gamma^{\epsilon}_{\mu\sigma} \
\overcirc{\Gamma}^{\alpha}_{\epsilon\alpha} -
\gamma^{\alpha}_{\epsilon\sigma} \
\overcirc{\Gamma}^{\epsilon}_{\mu\alpha}\\
&& - \ \gamma^{\alpha}_{\mu\epsilon} \
\overcirc{\Gamma}^{\epsilon}_{\sigma\alpha}) + \
R^{\epsilon}_{\sigma\alpha}(C^{\alpha}_{\mu\epsilon} - \
\overcirc{C}^{\alpha}_ {\mu\epsilon}) + \
\gamma^{\epsilon}_{\mu\sigma}\gamma^{\alpha}_{\epsilon\alpha} -
\gamma^{\epsilon}_{\mu\alpha}\gamma^{\alpha}_{\epsilon\sigma}.\\
\end{eqnarray*}
\\[- 1.45 cm] Consequently,
$$ \ \overcirc{R}_{\mu\sigma} = \beta_{\mu{o\atop |}\sigma} -
\gamma^{\alpha}_{\mu\sigma{o\atop |}\alpha} -
\beta_{\epsilon}\gamma^{\epsilon}_{\mu\sigma} +
\gamma^{\epsilon}_{\mu\alpha} \gamma^{\alpha}_{\epsilon\sigma} +
G^{\alpha}_{\mu\epsilon}R^{\epsilon}_{\alpha\sigma}. \ \ \Box$$

In view of Proposition 6.2 (e) and (f) and Proposition 6.5, we
obtain \vspace{-0.1cm}
\begin{cor} The following identities holds:
\begin{description}
\item[(a)] $(\beta_{\mu|\sigma} - \beta_{\mu{o\atop |}\sigma}) -
(\gamma^{\alpha}_{\mu\sigma|\alpha} -
\gamma^{\alpha}_{\mu\sigma{o\atop |}\alpha}) =
(\gamma^{\epsilon}_{\mu\alpha}\Omega^{\alpha}_{\sigma\epsilon} -
2\beta_{\epsilon}\gamma^{\epsilon}_{\mu\sigma})$

\item[(b)]  $(B_{\mu||\sigma} - B_{\mu{o\atop ||}\sigma}) -
(G^{\alpha}_{\mu\sigma||\alpha} - G^{\alpha}_{\mu\sigma{o\atop
||}\alpha}) = (G^{\epsilon}_{\mu\alpha}D^{\alpha}_{\sigma\epsilon} -
2B_{\epsilon}G^{\epsilon}_{\mu\sigma})$.

\end{description}
\end{cor}

The next two tables summarize the results obtained in this section,
where the contracted curvatures are expressed in terms of the
fundamental tensors.

\begin{center}{\bf Table 5 (a): Second rank curvature tensors}\\[0.3 cm]
\begin{tabular}
{|c|c|c|c|c|c|c|}\hline
&&\\
{}&{\bf Skew-symmetric}&{\bf Symmetric}
\\[0.2cm]\hline
&&\\
{\bf Dual} &\footnotesize{$\widetilde{R}_{[\mu\sigma]} =
\epsilon_{\sigma\mu} - L_{\sigma\mu} + M_{[\sigma\mu]} +
N_{[\sigma\mu]}$}&\footnotesize{$\widetilde{R}_{(\mu\sigma)} =
\theta_{\mu\sigma} + M_{(\mu\sigma)} - N_{(\mu\sigma)}$}
\\[0.2cm]\cline{2-3}%hline
&&\\
 \ \ &\footnotesize{$\widetilde{S}_{[\mu\sigma]} = \o
{\epsilon}_{\sigma\mu}$}&\footnotesize{$\widetilde{S}_{(\mu\sigma)}
= \o {\theta}_{\mu\sigma}$}
\\[0.2cm]\hline
&&\\

{\bf Symmetric} &\footnotesize{$\widehat{R}_{[\mu\sigma]} =
\frac{1}{2}\widetilde{R}_{[\mu\sigma]} +
\frac{1}{4}\eta_{\sigma\mu}$}&
\footnotesize{$\widehat{R}_{(\mu\sigma)} = \frac{1}{2}
\widetilde{R}_{(\mu\sigma)} + \frac{1}{4}\{h_{\mu\sigma} -
\omega_{\mu\sigma} - \sigma_{\mu\sigma}\}$}
\\[0.2cm]\cline{2-3}%\hline
&&\\
 \ \ &\footnotesize{$\widehat{S}_{[\mu\sigma]} =
\frac{1}{2}\widetilde{S}_{[\mu\sigma]} + \frac{1}{4}\o
{\eta}_{\sigma\mu}$}& \footnotesize{$\widehat{S}_{(\mu\sigma)} =
\frac{1}{2} \widetilde{S}_{(\mu\sigma)} + \frac{1}{4}\{\o
{h}_{\mu\sigma} - \o {\omega}_{\mu\sigma} - \o
{\sigma}_{\mu\sigma}\}$}
\\[0.2cm]\hline
&&\\
{\bf Riemannian} &\footnotesize{$\overcirc{R}_{[\mu\sigma]} =
\frac{1}{2}L_{\mu\sigma} - F_{[\mu\sigma]}$}&
\footnotesize{$\overcirc{R}_{(\mu\sigma)} = \theta_{\mu\sigma} -
\frac{1}{2}(\psi_{\mu\sigma}  - \phi_{\mu\sigma}) -
\omega_{\mu\sigma} + M_{(\mu\sigma)} - F_{(\mu\sigma)}$}
\\[0.2cm]\cline{2-3}%hline
&&\\
 \ \ &\footnotesize{$\overcirc{S}_{[\mu\sigma]} = 0$}&
\footnotesize{$\overcirc{S}_{(\mu\sigma)} = \o {\theta}_{\mu\sigma}
- \frac{1}{2}(\o {\psi}_{\mu\sigma}  - \o {\phi}_{\mu\sigma}) - \o
{\omega}_{\mu\sigma}$}
\\[0.2cm]\hline
\end{tabular}
\end{center}

\bigskip

\begin{center}{\bf{Table 5 (b): h- and v-scalar curvature tensors}}
\\[0.4 cm]

\begin{tabular}
{|c|c|c|}\hline
&&\\
{}&\bf{h-scalar curvature}&\bf{v-scalar curvature}
\\[0.4 cm]\hline
&&\\
{\bf Dual}&$\widetilde{R}^{\sigma}_{\sigma} = \theta +
T$&$\widetilde{S}^{\sigma}_{\sigma} = \o {\theta}$
\\[0.4 cm]\hline
&&\\
{\bf Symmetric}&$\widehat{R}^{\sigma}_{\sigma} = \frac{1}{2}(\theta
+ T) - \frac{1}{4}(3\omega + \sigma)$&
$\widehat{S}^{\sigma}_{\sigma} = \frac{1}{2}\o {\theta} -
\frac{1}{4}(3\o {\omega} + \o {\sigma})$

\\[0.4 cm]\hline
&&\\
{\bf Riemannian}&$\overcirc{R}^{\sigma}_{\sigma} = \theta -
\frac{1}{2}(\psi - \phi) - \omega + G$&
$\overcirc{S}^{\sigma}_{\sigma} = \o {\theta} - \frac{1}{2}(\o
{\psi} - \o {\phi}) - \o {\omega}$
\\[0.4 cm]\hline
\end{tabular}
\end{center}

\Section{The $W$-tensors}

The $W$-tensor was first defined by M. Wanas in 1975 \cite{aa} and
has been used by F. Mikhail and M. Wanas \cite{aaa} to construct a
geometric theory unifying gravity and electromagnetism. Recently,
two of the authors of this paper studied some of the properties of
this tensor in the context of the classical AP-space \cite{AMR}.

\begin{defn} Let $(M,\lambda)$ be a generalized AP-space.
For a given $d$-connection $D = (N^{\alpha}_{\beta},
\Gamma^{\alpha}_{\mu\nu}, C^{\alpha}_{\mu\nu})$, the horizontal
$W$-tensor ($hW$-tensor) $H^{\alpha}_{\mu\nu\sigma}$ is defined by
the formula
$${\lambda}_{\mu|\nu\sigma} - {\lambda}_{\mu|\sigma\nu} =
{\lambda}_{\epsilon}H^{\epsilon}_{\mu\nu\sigma},$$ whereas the
vertical $W$-tensor ($vW$-tensor) $V^{\alpha}_{\mu\nu\sigma}$ is
defined by the formula
$${\lambda}_{\mu||\nu\sigma} - {\lambda}_{\mu||\sigma\nu} =
{\lambda}_{\epsilon}V^{\epsilon}_{\mu\nu\sigma},$$ where \lq\lq\,\!
$|$\rq\rq\, and \lq\lq\,\! $||$\rq\rq\, are the horizontal and the
vertical covariant derivatives with respect to the connection $D$.
\end{defn}

We now carry out the task of calculating the different $W$-tensors.
 As opposed to the classical AP-space, which admits essentially one
$W$-tensor corresponding to the dual connection, we here have 4
distinct $W$-tensors: the horizontal and vertical $W$-tensors
corresponding to the dual $d$-connection, the horizontal $W$-tensor
corresponding to the symmetric $d$-connection and, finally, the
horizontal $W$-tensor corresponding to the Riemannian
$d$-connection. The remaining $W$-tensors coincide with the
corresponding curvature tensors.

\vspace{5pt}

It is to be noted that some of the expressions obtained for the
$W$-tensors are relatively more compact than those obtained for the
corresponding curvature tensors.

\begin{thm} The hW-tensor $\widetilde{H}^{\alpha}_{\mu\nu\sigma}$, the
vW-tensor $\widetilde{V}^{\alpha}_{\mu\nu\sigma}$, the hW-tensor
$\widehat{H}^{\alpha}_{\mu\nu\sigma}$ and the hW-tensor \
$\overcirc{H}^{\alpha}_{\mu\nu\sigma}$ corresponding to the dual,
symmetric and the Riemannian\linebreak $d$-connections respectively
can be expressed in the form:
\begin{description}
\item[(a)] $\widetilde{H}^{\alpha}_{\mu\nu\sigma} =
\Lambda^{\alpha}_{\sigma\nu|\mu} +
\Lambda^{\epsilon}_{\nu\sigma}\Lambda^{\alpha}_{\mu\epsilon} +
\frak{S}_{\mu, \nu, \sigma} L^{\alpha}_{\mu\sigma\nu}.$
\item[(b)]  $\widetilde{V}^{\alpha}_{\mu\nu\sigma} =
T^{\alpha}_{\sigma\nu||\mu} +
T^{\epsilon}_{\nu\sigma}T^{\alpha}_{\mu\epsilon}.$
\item[(c)] $\widehat {H}^{\alpha}_{\mu\nu\sigma} =
\frac{1}{2}(\Lambda^{\alpha}_{\mu\nu|\sigma} -
\Lambda^{\alpha}_{\mu\sigma|\nu}) +
\frac{1}{4}(\Lambda^{\epsilon}_{\mu\nu}\Lambda^{\alpha}_{\sigma\epsilon}
- \Lambda^{\epsilon}_{\mu\sigma}\Lambda^{\alpha}_{\nu\epsilon}) +
\frac{1}{2}(\Lambda^{\epsilon}_{\sigma\nu}\Lambda^{\alpha}_{\epsilon\mu}).$
\item[(d)] $\overcirc{H}^{\alpha}_{\mu\nu\sigma} = \gamma^{\alpha}_{\mu\nu|\sigma}
 - \gamma^{\alpha}_{\mu\sigma|\nu} +
\gamma^{\epsilon}_{\mu\sigma}\gamma^{\alpha}_{\epsilon\nu} -
\gamma^{\epsilon}_{\mu\nu}\gamma^{\alpha}_{\epsilon\sigma} +
\Lambda^{\epsilon}_{\nu\sigma}\gamma^{\alpha}_{\mu\epsilon}.$
\end{description}
\end{thm}

%\begin{proof]
\prof We prove \textbf{(a)} only. The proof of the other parts is
similar. We have
$${\lambda}_{\epsilon}\widetilde{H}^{\epsilon}_{\mu\nu\sigma} =
 {\lambda}_{\epsilon}
\widetilde{R}^{\epsilon}_{\mu\sigma\nu} +
{\lambda}_{\mu\widetilde{|}\epsilon}\widetilde{\Lambda}
^{\epsilon}_{\sigma\nu} +
{\lambda}_{\mu\widetilde{||}\epsilon}R^{\epsilon}_{\sigma\nu}.$$
Hence, taking into account Theorem 4.6 (a), we obtain
\begin{eqnarray*} \widetilde{H}^{\alpha}_{\mu\nu\sigma}&=& \widetilde{R}^{\alpha}_
{\mu\sigma\nu} + \undersym{\lambda}{i}^{\alpha} (\delta_{\epsilon} \
\undersym{\lambda}{i}_{\mu} - \undersym{\lambda}{i}_{\beta}
\Gamma^{\beta}_{\epsilon\mu})\widetilde{\Lambda}^{\epsilon}_{\sigma\nu}
+ \undersym{\lambda}{i}^{\alpha}(\dot{\partial_{\epsilon}} \
\undersym{\lambda}{i}_{\mu} -
\undersym{\lambda}{i}_{\beta}C^{\beta}_{\epsilon\mu})R^{\epsilon}_{\sigma\nu}\\
&=&\widetilde{R}^{\alpha}_{\mu\sigma\nu} +
\Lambda^{\epsilon}_{\nu\sigma}(\Gamma^{\alpha}_{\mu\epsilon} -
\Gamma^{\alpha}_{\epsilon\mu}) +
R^{\epsilon}_{\sigma\nu}(C^{\alpha}_{\mu\epsilon} -
C^{\alpha}_{\epsilon\mu})\\
&=&\Lambda^{\alpha}_{\sigma\nu|\mu} +
C^{\alpha}_{\epsilon\mu}R^{\epsilon}_{\sigma\nu} +
L^{\alpha}_{\sigma\nu\mu} + L^{\alpha}_{\nu\mu\sigma} +
\Lambda^{\epsilon}_{\nu\sigma}
\Lambda^{\alpha}_{\mu\epsilon} + T^{\alpha}_{\mu\epsilon}R^{\epsilon}_{\sigma\nu}\\
&=&  \Lambda^{\alpha}_{\sigma\nu|\mu} +
T^{\alpha}_{\epsilon\mu}R^{\epsilon}_{\sigma\nu} +
C^{\alpha}_{\mu\epsilon}R^{\epsilon}_{\sigma\nu} +
L^{\alpha}_{\sigma\nu\mu} + L^{\alpha}_{\nu\mu\sigma} +
\Lambda^{\epsilon}_{\nu\sigma} \Lambda^{\alpha}_{\mu\epsilon}
%\\&&
+ \ T^{\alpha}_{\mu\epsilon}R^{\epsilon}_{\sigma\nu}\\
&=& \Lambda^{\alpha}_{\sigma\nu|\mu} +
\Lambda^{\epsilon}_{\nu\sigma}\Lambda^{\alpha}_{\mu\epsilon} +
\frak{S}_{\mu, \nu, \sigma} L^{\alpha}_{\mu\sigma\nu}. \ \
\Box\end{eqnarray*}

\begin{prop} Let $\widetilde{\cal H}_{\nu\sigma} :=
\widetilde{H}^{\alpha}_{\alpha\nu\sigma}$, $\widehat{\cal
H}_{\nu\sigma} := \widehat{H}^{\alpha}_{\alpha\nu\sigma}$ and \
$\overcirc{\cal H}_{\nu\sigma}: = \
\overcirc{H}^{\alpha}_{\alpha\nu\sigma}$ with similar expression for
$\widetilde{\cal V}_{\nu\sigma}$. Then, we have
\begin{description}
\item[(a)]
$\widetilde {\cal {H}}_{\nu\sigma} = \beta_{\sigma|\nu} -
\beta_{\nu|\sigma} +
2\beta_{\epsilon}\Lambda^{\epsilon}_{\sigma\nu},$
\item[(b)] $\widetilde {\cal {V}}_{\nu\sigma} =
B_{\sigma||\nu} - B_{\nu||\sigma} +
2B_{\epsilon}T^{\epsilon}_{\sigma\nu},$
\item[(c)] $\widehat{\cal {H}}_{\nu\sigma} =
\frac{1}{2}\{\widetilde{\cal {H}}_{\nu\sigma} + \beta_{\epsilon}
\Lambda^{\epsilon}_{\nu\sigma}\},$
\item[(d)] $\overcirc{\cal {H}}_{\nu\sigma} = 0.$
\end{description}
\end{prop}

\begin{prop} Let $\widetilde{H}_{\mu\sigma} := \widetilde{H}^{\alpha}_{\mu\alpha\sigma}$,
$\widehat{H}_{\mu\sigma} := \widehat{H}^{\alpha}_{\mu\alpha\sigma}$
and \ $\overcirc{H}_{\mu\sigma} := \
\overcirc{H}^{\alpha}_{\mu\alpha\sigma}$ with similar expressions
for $\widetilde{V}_{\mu\sigma}$. Then, we have
\begin{description}
\item[(a)] $\widetilde{H}_{\mu\sigma} = \beta_{\sigma|\mu} +
\Lambda^{\epsilon}_{\alpha\sigma}\Lambda^{\alpha}_{\mu\epsilon} +
\frak{S}_{\alpha, \mu, \sigma} L^{\alpha}_{\alpha\mu\sigma},$

\item[(b)] $\widetilde{V}_{\mu\sigma} = B_{\sigma||\mu} + T^{\epsilon}_{\alpha\sigma}
T^{\alpha}_{\mu\epsilon},$

\item[(c)] $\widehat{H}_{\mu\sigma} = \frac{1}{2} \widetilde{H}_{\mu\sigma} +
\frac{1}{4}(\beta_{\epsilon}\Lambda^{\epsilon}_{\sigma\mu} +
\Lambda^{\epsilon}_{\sigma\alpha}\Lambda^{\alpha}_{\mu\epsilon}),$

\item[(d)] $\overcirc{H}_{\mu\sigma} = \beta_{\mu|\sigma} -
\gamma^{\alpha}_{\mu\sigma|\alpha} + \beta_{\epsilon}
\gamma^{\epsilon}_{\mu\sigma} - \gamma^{\epsilon}_{\sigma\alpha}
\gamma^{\alpha}_{\mu\epsilon}.$

\end{description}
\end{prop}

%In view of the above Proposition, we have
\begin{prop} The following holds:
\begin{description}
\item[(a)] $\widetilde{H}_{[\mu\sigma]} = \frac{1}{2}\{\beta_{\sigma|\mu} - \beta_{\mu|\sigma}\} +
\frak{S}_{\alpha, \mu, \sigma} L^{\alpha}_{\alpha\mu\sigma},$
\item[(b)] $\widetilde{H}_{(\mu\sigma)} = \frac{1}{2}\{\beta_{\sigma|\mu} + \beta_{\mu|\sigma}\} +
\Lambda^{\epsilon}_{\alpha\sigma}\Lambda^{\alpha}_{\mu\epsilon},$
\item[(c)] $\widetilde{V}_{[\mu\sigma]} = \frac{1}{2}\{B_{\sigma||\mu} - B_{\mu||\sigma}\},$
\item[(d)] $\widetilde{V}_{(\mu\sigma)} = \frac{1}{2}\{B_{\sigma||\mu} + B_{\mu||\sigma}\} +
T^{\epsilon}_{\alpha\sigma}T^{\alpha}_{\mu\epsilon},$
\item[(e)] $\widehat{H}_{[\mu\sigma]} = \frac{1}{2} \widetilde{H}_{[\mu\sigma]} + \frac{1}{4}
\beta_{\epsilon}\Lambda^{\epsilon}_{\sigma\mu},$
\item[(f)] $\widehat{H}_{(\mu\sigma)} = \frac{1}{2} \widetilde{H}_{(\mu\sigma)} + \frac{1}{4}
\Lambda^{\epsilon}_{\sigma\alpha}\Lambda^{\alpha}_{\mu\epsilon},$
\item [(g)] $\overcirc{H}_{[\mu\sigma]} = \frac{1}{2}\frak{S}_{\alpha\mu\sigma}
L^{\alpha}_{\alpha\mu\sigma},$
\item[(h)] $\overcirc{H}_{(\mu\sigma)} = \frac{1}{2}\{(\beta_{\mu|\sigma} + \beta_{\sigma|\mu}) -
\Omega^{\alpha}_{\mu\sigma|\alpha} +
\beta_{\epsilon}\Omega^{\epsilon}_{\mu\sigma}\} -
\gamma^{\alpha}_{\mu\epsilon}\gamma^{\epsilon}_{\sigma\alpha}.$
\end{description}
\end{prop}

%%%%%%%%%%%%%%%%%%%%%%%%%
%Taking the trace of the above tensors, we get
\begin{cor} the following holds:
\begin{description}
\item[(a)] $\widetilde{H}^{\alpha}_{\alpha} = \beta^{\alpha} \! _{|\alpha} +
\Lambda^{\epsilon\mu} \! _{\alpha}\Lambda^{\alpha}_{\epsilon\mu},$
\item[(b)] $\widetilde{V}^{\alpha}_{\alpha} = B^{\alpha} \! _{||\alpha} +
T^{\epsilon\mu} \! _{\alpha}T^{\alpha}_{\epsilon\mu},$
\item[(c)] $\widehat{H}^{\alpha}_{\alpha} = \frac{1}{2}\beta^{\alpha} \! _{|\alpha} +
\frac{1}{4} \Lambda^{\epsilon\mu} \!
_{\alpha}\Lambda^{\alpha}_{\epsilon\mu},$
\item[(d)] $\overcirc{H}^{\sigma} \!_{\sigma} = \beta^{\sigma} \!_{|\sigma} -
\frac{1}{2}\Omega^{\alpha\sigma} \!_{\sigma|\alpha} +
\frac{1}{2}\beta_{\alpha} \Omega^{\alpha\sigma} \!_{\sigma} -
\gamma^{\alpha\sigma} \!_{\epsilon}
\gamma^{\epsilon}_{\sigma\alpha}.$
\end{description}
\end{cor}
%%%%%%%%%%%%%%%%%%%%%%%%%%%%%%%%%%%%%%%%
Taking into account Proposition 4.4, Theorem 7.2 and the Bianchi
identity \cite{MM} for the Riemannian d-connection, we get  the
following
\begin{prop} The hW-tensors $\widetilde{H}^{\alpha}_{\mu\nu\sigma}$,
$\widehat {H}^{\alpha}_{\mu\nu\sigma}$, \
$\overcirc{H}^{\alpha}_{\mu\nu\sigma}$ and the vW-tensors
$\widetilde{V}^{\alpha}_{\mu\nu\sigma}$ satisfy the following
identities:
\begin{description}
\item[(a)] $\mathfrak{S}_{\mu, \nu, \sigma}\, \widetilde{H}^{\alpha}_{\mu\nu\sigma} =
2 \mathfrak{S}_{\mu, \nu, \sigma}(\Lambda^{\alpha}_{\mu\epsilon}\,
\Lambda^{\epsilon}_{\nu\sigma} + L^{\alpha}_{\mu\sigma\nu}).$
\item[(b)] $\mathfrak{S}_{\mu, \nu, \sigma}\, \widetilde{V}^{\alpha}_{\mu\nu\sigma} =
2 \mathfrak{S}_{\mu, \nu,
\sigma}(T^{\alpha}_{\mu\epsilon}\,T^{\epsilon}_{\nu\sigma}).$
\item [(c)] $\mathfrak{S}_{\mu, \nu, \sigma}\, \widehat{H}^{\alpha}_{\mu\nu\sigma} =
\mathfrak{S}_{\mu, \nu, \sigma}\,L^{\alpha}_{\mu\sigma\nu}.$
\item [(d)] $\mathfrak{S}_{\mu, \nu, \sigma} \ \overcirc{H}^{\alpha}_{\mu\nu\sigma} =
\mathfrak{S}_{\mu, \nu, \sigma}\,L^{\alpha}_{\mu\sigma\nu}.$
\end{description}
\end{prop}

We collect the results obtained in this section in the following
tables, where the contracted $W$-tensors are expressed in terms of
the fundamental tensors.
%The results obtained in Proposition 7.3, Proposition 7.5 and
%Corollary 7.6 are summarized in terms of the fundamental tensors in
%the following tables:
\vspace{0.5cm}
\begin{center}{\bf Table 6 (a): Second rank W-tensors}\\[0.3 cm]
\begin{tabular}
{|c|c|c|c|c|c|c|}\hline
&&\\
{}&{\bf Skew-symmetric}&{\bf Symmetric}
\\[0.2cm]\hline
&&\\
{\bf Dual} &\footnotesize{$\widetilde{H}_{[\mu\sigma]} =
\epsilon_{\sigma\mu} - L_{\sigma\mu} + 2M_{[\sigma\mu]}$}
&\footnotesize{$\widetilde{H}_{(\mu\sigma)} = \theta_{\mu\sigma} -
(\omega_{\mu\sigma} + \sigma_{\mu\sigma} - h_{\mu\sigma})$}
\\[0.2cm]\cline{2-3}
&&\\
\ \ &\footnotesize{$\widetilde{V}_{[\mu\sigma]} =
\bar{\epsilon}_{\sigma\mu}$}
&\footnotesize{$\widetilde{V}_{(\mu\sigma)} =
\bar{\theta}_{\mu\sigma} - (\bar{\omega}_{\mu\sigma} +
\bar{\sigma}_{\mu\sigma} - \bar{h}_{\mu\sigma})$}
\\[0.2cm]\hline
&&\\
{\bf Symmetric} &\footnotesize{$\widehat{H}_{[\mu\sigma]} =
\frac{1}{2}\widetilde{H}_{[\mu\sigma]} +
\frac{1}{4}\eta_{\sigma\mu}$}&
\footnotesize{$\widehat{H}_{(\mu\sigma)} = \frac{1}{2}
\widetilde{H}_{(\mu\sigma)} + \frac{1}{4}\{\omega_{\mu\sigma} +
\sigma_{\mu\sigma} - h_{\mu\sigma}\}$}
\\[0.2cm]\hline
&&\\
{\bf Riemannian}&\footnotesize{$\overcirc{H}_{[\mu\sigma]} =
\frac{1}{2}L_{\mu\sigma} - M_{[\mu\sigma]}$}&
\footnotesize{$\overcirc{H}_{(\mu\sigma)} = \theta_{\mu\sigma} -
\frac{1}{2}(\psi_{\mu\sigma}  - \phi_{\mu\sigma}) -
\omega_{\mu\sigma}$}
\\[0.2cm]\hline
\end{tabular}
\end{center}

\bigskip

\begin{center}{\bf{Table 6 (b): Scalar W-tensors}}

\bigskip

\begin{tabular}
{|c|c|c|}\hline
&&\\
&\bf{h-scalar W-tensors}&\bf{v-scalar W-tensors}
\\[0.4 cm]\hline
&&\\
{\bf Dual}&$\widetilde{H}^{\sigma}_{\sigma} = \theta - (3\omega +
\sigma)$& $\widetilde{V}^{\sigma}_{\sigma} = \bar{\theta} -
(3\bar{\omega} + \bar{\sigma})$
\\[0.4 cm]\hline
&&\\
{\bf Symmetric}&$\widehat{H}^{\sigma}_{\sigma} = \frac{1}{2}\theta -
\frac{1}{4}(3\omega + \sigma)$& \ \ %$\widehat{V}^{\sigma}_{\sigma} =
%\frac{1}{2}\bar{\theta} - \frac{1}{4}(3\bar{\omega} + \bar{\sigma})$

\\[0.4 cm]\hline
&&\\
{\bf Riemannian}&$\overcirc{H}^{\sigma}_{\sigma} = \theta -
\frac{1}{2}(\psi - \phi) - \omega$& \ \ %$\overcirc{V}^{\sigma}_{\sigma}
%= \bar{\theta} - \frac{1}{2}(\bar{\psi} - \bar{\phi}) -
%\bar{\omega}$
\\[0.4 cm]\hline
\end{tabular}
\end{center}

%%%%%%%%%%%%%%%%%%%%%%%%%%%%%%%%%%%%%%%%%%%%%%%%%%%%%%%%%%%%%% Concluding Remarks %%%%%%%%%%%%%%%%%%%%%%%%%%%%%%%%%%%%%%%%%%%%%%%%%%%%

\vspace{8pt}

 \noindent{\bf \Large {Concluding remarks}}

\vspace{12pt}

\par
 In the present article, we have
developed a parallelizable structure in the context of a generalized
Lagrange space. Four distinguished connections, depending on one
non-linear connection, are used to explore the properties of this
space. Different\linebreak curvature tensors characterizing this
structure are calculated. The contracted\linebreak curvature tensors
necessary for physical applications are given and compared (Tables
5(a)). The traces of these tensors are derived and compared (Table
5(b)). Finally, the different $W$-tensors
%, a fourth order tensor admitting both curvature and torsion,
with their contractions and traces are also derived (Tables 6(a) and
6(b)). \vspace{5pt}
\par
On the present work, we have the following comments and remarks:
\begin{enumerate}
\item Existing theories of gravity suffer from some problems connected to
recent\linebreak observed astrophysical phenomena, especially those
admitting {\bf anisotropic}\linebreak behavior of the system
concerned (e.g. the flatness of the rotation curves of spiral
galaxies). So, theories in which the gravitational potential depends
{\it on both position and direction} are needed. Such theories are
to be constructed in spaces admitting this dependence. This is one
of the aims motivating the present work.
\item Among the advantages of the AP-geometry are the ease in calculations and the
diverse and its thorough applications. In this work, we have kept as
close as possible to the classical AP-case. However, the extra
degrees of freedom in our GAP-geometry have created an abundance of
geometric objects which have no counterpart in the classical
AP-geometry. Since the physical meaning of most of the geometric
objects of the classical AP-structure is clear, we hope to attribute
physical meaning to the new geometric objects appearing in the
present work, especially the vertical quantities.

\item Due to the wealth of the GAP-geometry, one is faced with the problem of choosing
geometric objects that represent true physical quantities. As a
first step to solve this problem, we have written all second order
tensors in terms of the fundamental tensors defined in section 5.
This is done to facilitate comparison between these tensors and to
be able to choose the most appropriate  for physical application.
The same procedure has been used for scalars.

\item The paper is
not intended to be an end in itself. In it, we try to construct a
geometric framework capable of dealing with and describing physical
phenomena. The success of the classical AP-geometry in physical
applications made us choose this geometry as a guide line.

\par
The physical interpretation of the geometric objects existing in the
GAP-\linebreak geometry and not in the AP-geometry will be further
investigated in a forthcoming paper.

\end{enumerate}
\newpage
%%%%%%%%%%%%%%%%%%%%%%%%%%%%%%%%%%%%%%%%%%%%%%%%%%%%%% References %%%%%%%%%%%%%%%%%%%%%%%%%%%%%%%%%%%%%%%%%%%%%%%%%%%%%%%%%%%

\bibliographystyle{plain}

\end{document}